\documentclass[11pt]{article}
\pdfoutput=1

\usepackage{amsmath}
\numberwithin{equation}{section}
\numberwithin{figure}{section}
\numberwithin{table}{section}

\usepackage{mathtools}
\usepackage[inline]{enumitem}
\usepackage{booktabs}
\usepackage{graphicx}
\usepackage{amssymb}
\usepackage{amsthm}

\usepackage[numbers, sort&compress]{natbib}
\renewcommand\thesection{\arabic{section}}

\renewcommand\thefigure{\arabic{section}.\arabic{figure}}

\usepackage{color}
\usepackage[colorlinks]{hyperref}
\hypersetup{colorlinks=true, urlcolor=black, citecolor=black, runcolor=black, menucolor=black, filecolor=black, anchorcolor=black, linkcolor=black}

\usepackage[protrusion=true]{microtype}
\usepackage{breqn}
\usepackage{slashed}

\newcommand{\dagg}{^{\dagger}}

\newcommand{\mL}{\mathcal{L}}

\newcommand{\mO}{\mathcal{O}}
\newcommand{\mM}{\mathcal{M}}

\newcommand{\set}[1]{\left\{#1\right\}}

\newcommand{\SU}[1]{\textrm{SU}\left(#1\right)}

\newcommand{\Uni}[1]{\textrm{U}\left(#1\right)}

\newcommand{\upd}[1]{\text{d}#1 \,}
\newcommand{\ud}{\text{d}}

\newcommand{\thrbythr}[3]{\begin{pmatrix}#1 \\ #2 \\ #3\end{pmatrix}}
\newcount\colveccount
\newcommand*\colvec[1]{
        \global\colveccount#1
        \begin{pmatrix}
        \colvecnext
}
\def\colvecnext#1{
        #1
        \global\advance\colveccount-1
        \ifnum\colveccount>0
                \\
                \expandafter\colvecnext
        \else
                \end{pmatrix}
        \fi
}
\newcommand{\Abs}[1]{\left| #1 \right|}

\usepackage{jcappub}

\title{Linking Scalar Dark Matter and Neutrino Masses with IceCube 170922A}
\author[ ]{J.B.G. Alvey\note{\href{mailto:james.alvey@kcl.ac.uk}{james.alvey@kcl.ac.uk}}}
\author[ ]{and M. Fairbairn\note{\href{mailto:malcolm.fairbairn@kcl.ac.uk}{malcolm.fairbarin@kcl.ac.uk}}}
\affiliation[ ]{Theoretical Particle Physics and Cosmology \\ \emph{Department of Physics, King's College London, London WC2R 2LS, United Kingdom}}

\abstract{Two of the key unresolved issues facing Standard Model physics are (i) the appearance of a small but non-zero neutrino mass, and, (ii) the missing mass problem in the Universe. The focus of this paper is a previously proposed low energy effective theory that couples a dark scalar to Standard Model neutrinos. This provides a stable dark matter candidate as well as radiatively generating a neutrino mass. Within this framework we will then construct an entirely new bound from the IceCube-170922A event which takes into account (i) the possible neutrino mass hierarchies, (ii) the effect of cosmological redshift on e.g. the number density of cosmic neutrino background neutrinos, and, (iii) the non-degeneracy of neutrino mass and flavour eigenstates. This builds on work by Kelly and Machado (2018), where the authors placed new constraints on neutrinophilic and axion dark matter models. At low mediator masses, we find an improvement of an order of magnitude on current constraints from kaon decays.  The constraint is complimentary (and slightly weaker) than current constraints from Big Bang Nucleosynthesis and the Cosmic Microwave Background. We explore how future higher energy events could improve this bound.}
\usepackage[margin=1.0in]{geometry}

\renewcommand*{\thefootnote}{\fnsymbol{footnote}}
\begin{document}
\begin{flushright}
KCL-PH-TH-2019-013
\end{flushright}
\maketitle
\flushbottom
\newpage
\renewcommand{\thefootnote}{\arabic{footnote}}
\section{Introduction}

There are two glaring issues with the Standard Model of particle physics \cite{Bonnet2012, Farzan2011, Bednyakov2007, Kubo2006, Davidson2002, Ma2001, Yao2018}. Firstly, the Standard Model neutrino does not have a mass, but nonetheless observations of neutrino oscillations imply the opposite. Secondly, there is strong evidence to suggest that there is a missing mass problem in the Universe.  While we are agnostic as to the explanation for this problem, the fact is that it is neatly solved by particle dark matter.  In this work, we are interested in constraining effective models of particle Dark Matter that generate neutrino masses consistently. Note that this suggests a coupling between the Dark Matter and the Standard Model neutrinos. In particular we are looking to constrain a subset of interactions as detailed in Table I of \cite{BoehmPascoli} concerning scalar dark matter. In the paper by Kelly and Machado \cite{Kelly}, constraints were placed on the vector models also using data from IceCube.

The model of scalar Dark Matter we are considering is proposed in \cite{Boehm,Farzan2009, Ma2006, Franarin2018, Farzan2010, Artamonov2016, Farzan2010a, Farzan2011, Ambrosino2009, Farzan2014, Ma2006a, Boehm2006, Serpico2004, Boehm2004, Boehm2003, Ma1998} and contains a coupling to the standard model neutrinos via a fermionic mediator. The mass range being considered is $\mO(\textrm{MeV})$. There are number of constraints placed on this model due to relic abundance, kaon decay, and neutrino mass considerations. Particular attention is paid to the constraints on the scalar mass from observations of the Cosmic Microwave Background (CMB) and light element abundances. The centre of mass energy of the $\nu\nu$ interaction under consideration means that our bounds lie in a similar region of parameter space. A discussion of this matter as well as the outlook for observing higher energy neutrinos is given in Section \ref{sec:cmbconstraints}.

We are interested in computing complementary bounds using the IceCube neutrino experiment. IceCube is a relatively new facility located at the South Pole, which detects high energy astrophysical neutrinos \cite{Kelly, Padovani2018, IceCube, IceCube2018, Ackermann2018, Difranzo2015, Hooper2018, Ioka2014}. On the $22^{\textrm{nd}}$ September $2017$, IceCube detected a $290\,\textrm{TeV}$ muon neutrino. After analysis of the trajectory of the event, it is generally believed that this neutrino came from the blazar TXS $0506$+$056$. This active galactic nuclei is located approximately $1.3 \,\textrm{Gpc}$ away in comoving distance co-ordinates.

Two facts are key to our analysis. Firstly the centre of mass energy of the high energy neutrino interacting with the cosmic neutrino background is at the MeV scale. The fact that this is of the same order as the scalar mass under consideration leads to an enhancement of the cross-section for the process $\nu\nu \rightarrow \delta\delta$. In turn, this reduces the mean free path to a value well below the Standard Model equivalent. Here, $\delta$ is the scalar dark matter candidate. Secondly, the fact that the blazar neutrino can propagate over gigaparsecs of cosmological distance means that we may be able to compute the mean free path of the neutrino within the new framework and find something comparable with this distance. Note that this is non-trivial in the sense that the Standard Model prediction for the neutrino mean free path after the neutrino sector decouples from the photon bath is $\mO(10^{11}) \, \textrm{Gpc}$.

In order to make this comparison, we need to include physics within the neutrino sector, as well as astrophysical and cosmological considerations \cite{Bertuzzo, Bellini2018, Ringwald2004, Couchot2017, King, dermer2009high, Karmakar2018, Pandey2019}. In the neutrino case, we wish to include the phenomenology of the neutrino mass hierarchy, as well as the non-degeneracy of the neutrino mass and flavour eigenstates. On astrophysical grounds, it is not obvious that we should be allowed to compare the mean free path of the neutrinos and the distance to the blazar to make a deduction. Blazar neutrino production mechanisms are somewhat uncertain \cite{Padovani2018, Padovani2019, Keivani2018}, so we must be careful to consider which physics is essentially model independent, and use this to support our conclusions regarding the mean free path. This will involve a multimessenger approach across neutrino experiments and $\gamma$-ray telescopes. We should note at this point that this methodology is not novel, specifically a clear exposition of the ideas is given in \cite{Kelly}. In this case, the authors make use of the lower energy events from the previous flaring period as opposed to the $290$ TeV event we consider here. There are some additional subtleties that arise because of the higher energy that we address in Section \ref{sec:nuluminosity}. Finally, cosmological redshift affects the number densities of the relic neutrinos that act as the scattering medium.

The result of the study is that we find competitive bounds arising from our analysis. Indeed in the low mass regime for the scalar Dark Matter and mediator, we find an improvement of an order of magnitude in the case that the Dark Matter is complex. We were unable to say anything in the real case as the couplings are already too tightly constrained.

The outline of the paper is as follows. In Section \ref{sec:dmneutrinos}, we discuss the generation of neutrino mass beyond the Standard Model, and the importance of a residual $\mathbb{Z}_2$ symmetry in stabilising the Dark Matter candidate. In Section \ref{sec:low energy theory}, we present the effective field theory for the model, explaining the particle content. More importantly, we also explain the current constraints on the model, differentiating between the real and complex scenarios. In Sections \ref{sec:icecube} and \ref{sec:nuluminosity}, we introduce the IceCube experiment and the blazar TXS $0506$+$056$, and present the key calculation that motivates our usage of the mean free path as a comparison. Section \ref{sec:assumptions} considers the details regarding the neutrino mass hierarchy, flavour structure, redshift effects, and the assumption that the other possible processes are subdominant. We present our results in Section \ref{sec:results} along with an analysis. Finally, in Section \ref{sec:conclusion}, we draw some conclusions regarding the approach.

\section{Linking Dark Matter and Neutrino Mass}\label{sec:dmneutrinos}

We begin the paper by providing a motivation for linking the two issues facing the Standard Model. In particular, we discuss the different possibilities for generating neutrino mass. We also introduce the crucial role that a $\mathbb{Z}_2$ symmetry plays in both stabilising the Dark Matter candidate, and suppressing the neutrino mass contribution at tree level.

\subsection{How to Generate a Neutrino Mass}

In the neutrino sector, there are two possible types of mass term in the Lagrangian \cite{Ma1998, Bellini2018, Boehm2006, Yao2018};
\begin{itemize}
  \item \textit{Dirac Neutrino Masses} arise after the introduction of a right-handed neutrino field $\nu^i_R$ with $i = e, \mu, \tau$. The mass then arises under the Higgs mechanism due to a coupling;
  \begin{equation}
    \mL_{\textrm{\small lept}, \phi} \propto \lambda^{ij}\bar{\psi}^i\Phi\ell^j + \lambda_\nu^{ij}\bar{\psi}^i\Phi^c \nu^i_R
  \end{equation}
  where $\psi^i$ is the left handed $\SU{2}$ lepton doublet, $\Phi$ is the $\SU{2}$ Higgs doublet, and $\ell^j$ is the right-handed lepton singlet. After electroweak symmetry breaking (EWSB), the neutrinos get a mass term;
  \begin{equation}
    -\sum_{i}{m_\nu^i \left(\bar{\nu}^i_R \nu^i_L + \bar{\nu}^i_L \nu^i_R\right)}
  \end{equation}
  \item \textit{Majorana Masses} are a qualitatively different scenario that is possible if the fermion is neutral, as in the case of the neutrino. In this case, the right-handed neutrino field is not independent of the left-handed one.\footnote{To be precise, $\nu_R(x) =
  \nu_L^c(x)$ where $c$ represents the charge conjugated field.} The mass term becomes;
  \begin{equation}\label{eq:Majorana Mass Term}
    -\frac{1}{2}\sum_{i}{m_\nu^i \left(\bar{\nu}_L^{i,c}\nu_L^i + \bar{\nu}^i_L \nu^{i, c}_L\right)}
  \end{equation}
\end{itemize}
We will focus on the second scenario. This can't arise at tree level in the Standard Model. The lowest dimension for which such a term is generated is at dimension $5$ via an operator of the form \cite{Bonnet2012};
\begin{equation}
  \Lambda^{-1}\phi^0 \phi^0 \nu^i_L \nu^j_L
\end{equation}
Being dimension $5$, the operator is not renormalisable, and as such this can only be an effective coupling, valid up to some large mass scale $\Lambda$.

\subsection{Dark Matter and the Role of $\mathbb{Z}_2$}\label{sec:understanding Z2}

The fact that approximately $26$\% \cite{Planck} of the universe at the current time is formed of non-baryonic matter generates a number of questions about the nature of this hidden sector. These include (i) what, if any, are the couplings to the Standard Model? and, (ii) what evidence can be used to distinguish and constrain different models? One of the key properties of a particle Dark Matter candidate is that it should be stable. In the effective theory case, we will claim that our Dark Matter candidate is stable ``as a result of a $\mathbb{Z}_2$ symmetry'' \cite{Ma2006a, Kubo2006, Ma2006, Ma2001}.

The big picture is as follows; suppose a new scalar is introduced into the Standard Model which is odd under a new global $\mathbb{Z}_2$ symmetry. Through interactions with Standard Model neutrinos and new heavy fermions, it will generate a very similar dimension $5$ operator in a completely analogous way to the neutral Higgs field $\phi^0$. Unlike the Higgs however, the effect of the $\mathbb{Z}_2$ symmetry on the Higgs potential ensures that the new scalar cannot develop a VEV. It also means that the lightest degree of freedom is rendered stable. These two facts have the combined effect that (i) the neutrino mass is still generated in this process, but at one loop instead of at tree level, (ii) the lightest degree of freedom in the scalar sector is stable and can therefore act as our Dark Matter candidate.

For a review of the original seesaw mechanism, see \cite{Ma1998,Ma2001,Ma2006a}. Also, in \cite{Ma2006a} the argument is presented in the framework of the $\mathbb{Z}_2$ symmetry and explicitly identifies the candidate in the scalar spectrum. Finally, to understand the key fact that the neutrino mass is generated \emph{only} at one loop, one should consult \cite{Ma2006}.

\section{The Low Energy Effective Field Theory}\label{sec:low energy theory}

The discussion of UV complete theories \cite{Farzan2009, Farzan2010, Farzan2010a} presents a concrete scenario to implement the ideas of radiatively generating neutrino mass, as well as providing a dark matter candidate. On the other hand, at a given low energy scale, our experiments will not be able to probe the UV structure of the theory, only the effective degrees of freedom remaining at the scale in question (e.g. the energy scale at the LHC, or in cosmological scenarios).

In this section, we will present a low energy effective theory that contains a scalar Dark Matter candidate coupled to neutrinos. This was originally proposed in \cite{Farzan2014, Farzan2011, Farzan2010, Boehm2006, Farzan2009}. A key distinction will be between real and complex scalar Dark Matter, since the two have qualitatively different constraints. We shall explain why it must be an effective theory, as well as indicate how the neutrino mass is generated. More importantly, we will focus on the phenomenological consequences of the model and be as clear as possible as to exactly what constraints these place on the couplings and masses within the theory. These include astrophsyical, particle physics, and cosmological bounds \cite{Farzan2014, Farzan2011, Farzan2010, Boehm2006, Farzan2009}. This will provide the basis for further numerical investigation.

It is not within the scope of this work to present a full UV completion of this effective theory. Nonetheless, \cite{Farzan2009, Farzan2010} present such a completion in the real case. In \cite{Farzan2010a} the same is done for complex dark matter. The fact that the particle content of the effective model can arise from a fully gauge invariant theory is important since it provides a clear understanding of where the relevant degrees of freedom come from, as well as the scale of new physics.

\subsection{The Lagrangian and Particle Content}

Our focus is on a low energy particle spectrum that consists of the Standard Model along with \cite{Farzan2014, Farzan2011};
\begin{itemize}
  \item A scalar field, $\delta$, which may be real or complex;
  \item Two or more massive right-handed fermions $N^i_R$, in what follows, we will assume these to be of \textit{Majorana} type.
\end{itemize}
In addition to this, we assume that there is a residual $\mathbb{Z}_2$ symmetry \cite{Ma1998, Ma2001, Ma2006} under which the new particles are odd, and the Standard Model is even. This has a number of effects, in a fashion completely analagous to the discussion in Section \ref{sec:understanding Z2};
\begin{enumerate}
  \item It ensures that the lightest particle in the spectrum is stable, providing a Dark Matter candidate
  \item It prohibits a term of the form $\phi^0 \bar{N}_R^i \nu^\alpha_L$, so that after EWSB, there is no Dirac mass term linking $N^i_R$ and $\nu^\alpha_L$. This means that the neutrino mass is not generated at tree level.
  \item The new scalar, $\delta$, cannot acquire a VEV, due to the structure of the Higgs sector
\end{enumerate}
The key feature we are interested in however is the interaction part of the Lagrangian which couples this new dark sector to the neutrino sector in the Standard Model. In our effective theory, we take the couplings to be of the form;
\begin{equation}\label{eq:effective lagrangian}
  g_{i\alpha}\delta \bar{N}^i_R \nu^\alpha_L + \text{h.c.}
\end{equation}
\noindent Where there is an implicit sum over the right-handed fermion species, $i$, and the Standard Model neutrino species, $\alpha$. Later we will be interested in constraining the values of the coupling constants.
\subsubsection{Why must it be an effective theory?}\label{sec:why effective}
To understand why the contribution to the Lagrangian in \eqref{eq:effective lagrangian} must be effective \cite{Farzan2009, Farzan2010a}, i.e. only valid up to some high energy scale $\Lambda$, we make an assumption regarding the representations of the new fields. We assume that there are no electroweak interactions between $\delta$, $N^i_R$ and the electroweak gauge bosons. In other words, they lie in the trivial representation of $\SU{2}_L \times \Uni{1}_Y$. This is important because the Standard Model neutrinos are charged under the electroweak symmetry. Therefore the coupling in \eqref{eq:effective lagrangian} cannot be gauge invariant. It must therefore be an effective theory that is part of some $\SU{2}_L \times \Uni{1}_Y$ invariant UV theory. The implication of this is that when calculating for example loop diagrams, we should bear in mind that the theory does not hold up to arbitrarily high energies. Instead there is some cutoff, $\Lambda$, below which the Lagrangian is a valid description.

\subsection{Generation of Neutrino Masses}

In this section we will present the contribution to the neutrino mass matrix that arises due to interaction term in \eqref{eq:effective lagrangian}. The Majorana mass term for the neutrino is of the form;
\begin{equation}\label{eq:neutrino mass}
  \frac{1}{2}(m_\nu)_{\alpha\beta}\left(\bar{\nu}^\alpha_L \nu^\beta_L + \text{h.c.}\right)
\end{equation}
We are interested in the mass matrix $(m_\nu)_{\alpha\beta}$. To do so, we will need to distinguish between the two cases where either (i) $\delta$ is a real scalar field, or, (ii) $\delta$ is a complex scalar field.
\subsubsection{Case 1: Real Scalar Field}
\begin{figure}
  \centering
  \includegraphics[width=0.7\linewidth]{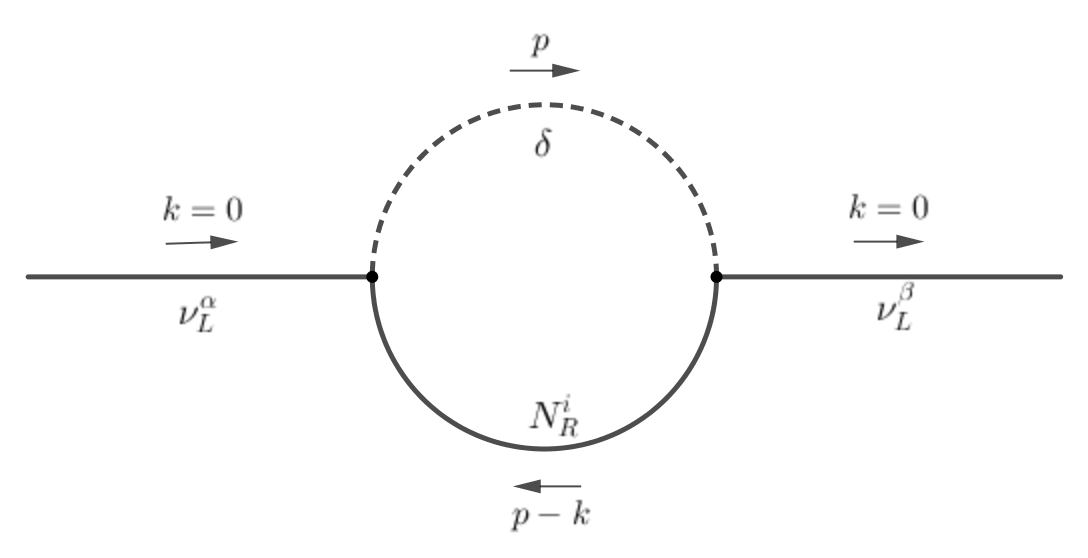}
  \caption{The one loop diagram contributing to the neutrino mass in the case that $\delta$ is a real scalar. The external neutrinos are evaluated at zero incoming momenta so as to extract only the mass contribution as opposed to the quadratic derivative interactions.}
  \label{fig:onelooprealdiag}
\end{figure}

In this case, there is only one diagram that contributes to the neutrino mass, as shown in Figure \ref{fig:onelooprealdiag}. As in \cite{Farzan2010, Boehm2006, Farzan2011}, the result in the real case is;
\begin{equation}\label{eq:oneloopmass result}
  (m_\nu)_{\alpha\beta} = \sum_{i}{\frac{g_{i\alpha}g_{i\beta}}{16\pi^2}m_{N^i}\left(\log\frac{\Lambda^2}{m_{N^i}^2} - \frac{m_\delta^2}{m_{N^i}^2 - m_\delta^2}\log\frac{m_{N^i}^2}{m_\delta^2}\right)}
\end{equation}

\subsubsection{Case 2: Complex Scalar Field}

To repeat the calculation for the complex scalar field, we should first think about the scalar degrees of freedom. Since the field is in the trivial representation of the electroweak gauge group, the most general hermitian mass term for $\delta$ can be written as \cite{Farzan2010};
\begin{equation}
  V_m = M^2 \delta\dagg \delta - \frac{1}{2}(m^2 \delta\delta + \text{h.c.})
\end{equation}
Consider expanding the scalar field as $\delta = \tfrac{1}{\sqrt{2}}(\delta_1 + i\delta_2)$ where $\delta_{1,2}$ are both real fields. Expanding the terms above in terms of these real degrees of freedom;
\begin{dmath}
  V_m = \frac{1}{2}M^2(\delta_1 + i\delta_2)(\delta_1 - i\delta_2) - \frac{1}{4}m^2\left((\delta_1 + i\delta_2)(\delta_1 + i\delta_2) + (\delta_1 - i\delta_2)(\delta_1 - i\delta_2)\right)
\end{dmath}
Collecting the terms together for each field we find;
\begin{equation}
  V_m = \frac{1}{2}(M^2 - m^2)\delta_1^2 + \frac{1}{2}(M^2 + m^2)\delta_2^2
\end{equation}
We can then immediately see that the mass eigenstates are simply $\delta_1$ and $\delta_2$ themselves, with masses $m^2_{\delta_1} = M^2 - m^2$, $m^2_{\delta_2} = M^2 + m^2$. The lighter of these will be our Dark Matter candidate. Also note that interaction term \eqref{eq:effective lagrangian} is diagonal in this mass basis;
\begin{equation}
  \mL_{\text{int}} = g_{i\alpha}\bar{N}^i_R \nu^\alpha_L (\delta_1 + i \delta_2)
\end{equation}
The contributions to the neutrino mass are then two diagrams of the form in Figure \ref{fig:onelooprealdiag}. One will have $\delta_1$ running round the loop, whilst the other will have $\delta_2$. We note that the only difference between them is that the second diagram will have two couplings with an extra factor of $i$, $(ig_{i\alpha})(ig_{i\beta}) = -g_{i\alpha}g_{i\beta}$. As such the second will come with a negative sign. The total contribution is then a sum of two contributions of the form \eqref{eq:oneloopmass result}. Importantly, we see that the dependence on the cutoff drops out in the complex case and we find;
\begin{equation}
  (m_\nu)_{\alpha\beta} = \sum_{i}{\frac{g_{i\alpha}g_{i\beta}}{16\pi^2}m_{N^i}\left(\frac{m_{\delta_2}^2}{m_{N^i}^2 - m_{\delta_2}^2}\log\frac{m_{N^i}^2}{m_{\delta_2}^2} - \frac{m_{\delta_1}^2}{m_{N^i}^2 - m_{\delta_1}^2}\log\frac{m_{N^i}^2}{m_{\delta_1}^2}\right)}
\end{equation}

\subsection{Astrophysical, Cosmological, and Particle Physics Constraints}

So far we have illustrated (i) how neutrino mass is generated within the effective framework, and, (ii) discussed which are the relevant degrees of freedom when it comes to suitable dark matter candidates. We now discuss the vital question as to what constraints have already been placed on the model across a range of scenarios. We will investigate the constraints that arise from the two key scenarios \cite{Boehm2006, Franarin2018, Farzan2009, Farzan2011, Farzan2014, Farzan2010};
\begin{enumerate}
  \item The cosmological bound on the Dark Matter annihilation cross section;
  \item The bounds that arise due to light meson and tau decay.
  \item Bounds that arise from observations of the CMB and Big Bang Nucleosynthesis (BBN)
\end{enumerate}
For completeness, we will also discuss some of the other astrophysical bounds in slightly more general terms.
\subsubsection{Dark Matter Annihilation Cross Section}
Within this effective model, there are three annihilation channels in the case that $N^i_R$ is a Majorana fermion, given by $\delta \delta \rightarrow \set{\nu\nu, \nu\bar{\nu}, \bar{\nu}\bar{\nu}}$. As in \cite{Boehm2006}, we take $\Lambda \sim 200 \, \text{GeV}$, $0.01 \, \text{eV} < m_\nu < 1\, \text{eV}$. Furthermore, we also use the fact that sub-GeV Dark Matter requires $\langle \sigma v \rangle \simeq 5 \times 10^{-26}\, \text{cm}^3\text{s}^{-1}$ in order for the correct relic abundance to be obtained \cite{Steigman2012}.
\subsubsection*{Real Scalar Dark Matter}
In the real case, this gives the constraints;
\begin{equation}
  \mO(1 \, \text{MeV}) \lesssim m_\delta < m_N \lesssim 10 \, \text{MeV}, \quad 3 \times 10^{-4} \lesssim g \lesssim 10^{-3}
\end{equation}
\noindent This is a strong bound on the coupling, and in fact we can \emph{only} obtain new constraints in the complex case. The real case is too weakly coupled. From now on therefore, whilst we will of course mention the real case, our focus will be on the complex scenario.

\subsubsection*{Complex Scalar Dark Matter}
In the complex case \cite{Boehm2006}, we get less stringent constraints on the masses;
  \begin{equation}
    (1\, \text{MeV})^2 \lesssim \Abs{m_{\delta_2}^2 - m_{\delta_1}^2} \lesssim (20 \, \text{MeV})^2
  \end{equation}
\noindent We note that now $m_N$ is a free parameter, so is far less constrained than in the real case. In turn, this implies that $g$ is less constrained, although we shall see that there are different bounds due to light meson decay.

\subsubsection{Light Meson and Tau Decay}

Consider the decay $K^+ \rightarrow e^+/\mu^+ + \nu$ \cite{Farzan2011, Farzan2014, Farzan2010}. Then generically if the coupling in \eqref{eq:effective lagrangian} is present, new decay modes $K^+ \rightarrow e^+/\mu^+ + N^i_R + \delta$ should exist. This means we should see $K^+ \rightarrow e^+/\mu^+ + \text{missing energy}$ compared to the Standard Model. This can be tested by experiments such as KLOE. Importantly, in this case the bounds are valid in the real \emph{and} complex scenarios. They are presented in \cite{Farzan2011, Farzan2014, Farzan2010, Artamonov2016};
  \begin{equation}
    \sum_{i}{\Abs{g_{ie}}^2} < 10^{-5}, \quad \sum_{i}{\Abs{g_{i\mu}}^2} \lesssim 10^{-4}, \quad \sum_{i}{\Abs{g_{i\tau}}^2} < 10^{-1}
  \end{equation}
\noindent There are a couple of important observations that need to be made with respect to these bounds;
\begin{itemize}
  \item Provided $\max(m_\delta, m_{N^i}) \ll m_{K, \pi} \simeq \mO(500\,\text{MeV})$, the bounds are similar in both the complex and the real case;
  \item With the bounds as above, we note that the couplings themselves can have magnitudes;
  \begin{equation}
    \label{eq:constraints}
    \Abs{g_{ie}} \lesssim 3 \times 10^{-3}, \quad \Abs{g_{i\mu}} \lesssim 10^{-2}, \quad \Abs{g_{i\tau}} \lesssim 3 \times 10^{-1}
  \end{equation}
  \item In the heavy case \cite{Farzan2014}, $m_K < m_\delta + m_{N^i} < m_D$, where $m_D$ is the mass of the $D$ meson, the bounds above do not apply. Instead the strongest bounds have $\Abs{g_{ie}} \lesssim 0.4$ and $g_{i\mu} \lesssim \mO(1)$.
\end{itemize}

\subsubsection{CMB Constraints on the Scalar Mass}\label{sec:cmbconstraints}

There are additional constraints on light particle dark matter that arise from measurements of the CMB, the light element abundance, and the expansion rate of the Universe. They apply here because, firstly, the existence of a light species such as our dark matter candidate during BBN at $z \sim 3 \times 10^8$ could change the rate of expansion and hence the time of weak freeze-out, changing relic abundances. Secondly, additional energy might be injected into the neutrino sector as this dark matter candidate annihilates. The relevant constraints are shown in Table \ref{tab:cmb} and are based on \cite{Boehm}, \cite{Nollett2015}, and \cite{Escudero2019}.

\begin{table}[b]
\centering
\begin{tabular}{lcc}
\toprule \textbf{Constraint} & \textbf{Mass Bound [MeV]} & \textbf{Min. Neutrino Energy [PeV]} \\
\midrule 
$N_{\textrm{\small eff}}$ \cite{Boehm} & 3.90 & 0.76 \\
BBN + Planck + $N_{\textrm{\small eff}}$ + $Y_p$ \cite{Nollett2015} & 6.74 & 2.27 \\
BBN + Planck + $N_{\textrm{\small eff}}$ \cite{Nollett2015} & 6.98 & 2.43 \\
Planck + BAO + $H_0$ + $N_{\textrm{\small eff}}$ \cite{Escudero2019} & 7.80 & 3.04 \\
\bottomrule
\end{tabular}
\caption{Table illustrating the constraints on complex scalar dark matter coming from various cosmological sources as well as the corresponding minimum neutrino energy that would need to be observed to probe parameter space outside of these bounds.}
\label{tab:cmb}
\end{table}

Kinematically, the constraints are relevant in this scenario because the centre of mass energy of the $\nu\nu \rightarrow \delta\delta$ interaction determines the maximum mass $m_\delta$ of the scalar particle that can be produced;
\begin{equation}
    m_\delta \leq \frac{1}{2}\sqrt{s} = \frac{1}{2}\sqrt{2 E_\nu m_\nu}
\end{equation}
where $E_\nu$ is the energy of the high-energy neutrino from the blazar. We see therefore that the CMB bounds on the mass of the scalar are related to a minimum neutrino energy above which the method set out in this paper will probe parameter space unconstrained by the CMB, although we note this is not quite the case for the currently observed event. Stated another way the fact we are only sensitive to MeV scale dark matter is partly due to the centre of mass energy for the $\nu\nu \rightarrow \delta\delta$ interaction being less than the threshold energy for larger scalar masses.  The $290$ TeV neutrino is not at the upper end of the expected blazar neutrino flux distribution, so it is not unreasonable to expect that higher energy neutrinos will be observed, immediately allowing us to probe higher mass regimes, less sensitive to CMB constraints. To investigate this quantitatively, we choose $m_\nu = 0.04\,\mathrm{eV}$ to be the lightest neutrino species. This ensures that $\sum{m_\nu} < 0.17 \, \mathrm{eV}$. Then, for each bound we may compute the minimum neutrino energy that must be observed to probe new parameter space. These reference neutrino energies are also shown in Table \ref{tab:cmb}. We consider how this maps onto the expected constraints that would be obtained with observations of higher energy neutrinos in Figure \ref{fig:mpmn}.

As a final comment, there is work being done currently that re-explores some of the assumptions in deriving these constraints based on entropic or decay arguments, see for example \cite{Kreisch}. Figure 3 in \cite{Wilkinson} proves a useful reference to see the separate constraints from BBN and the CMB. With this in mind, we view the bounds presented in this paper as complementary to those from Cosmology, requiring different (in this case astrophysical) assumptions.

\subsubsection{Additional Constraints}

The two methods of constraining the effective theory given above provide the most stringent bounds on the couplings and the masses. Furthermore, they also provide two very distinct scenarios for comparison. The first investigates the phenomenology in the setting of thermalizing early universe cosmology, whilst the second is a pure particle physics test. Within the literature \cite{Farzan2010, Boehm2006}, there are a couple of other suggestions for additional, or future methods of constraint. These include;
\begin{enumerate}
  \item \textit{Supernova Core Collapse:} It is suggested \cite{Franarin2018, Farzan2010} that we could search for a dip in the neutrino energy spectrum coming from neutrinos produced during supernova core collapse.
  \item \textit{Large Scale Structure:} This places constraints on the mass \cite{Boehm2004} due to the damping length of $\mO(\text{keV}) \lesssim m$.
\end{enumerate}

\subsection{$\nu\nu \rightarrow \delta\delta$ Cross Section}\label{sec:crosssection}

\begin{figure}[t]
\centering
\includegraphics[width=0.6\linewidth]{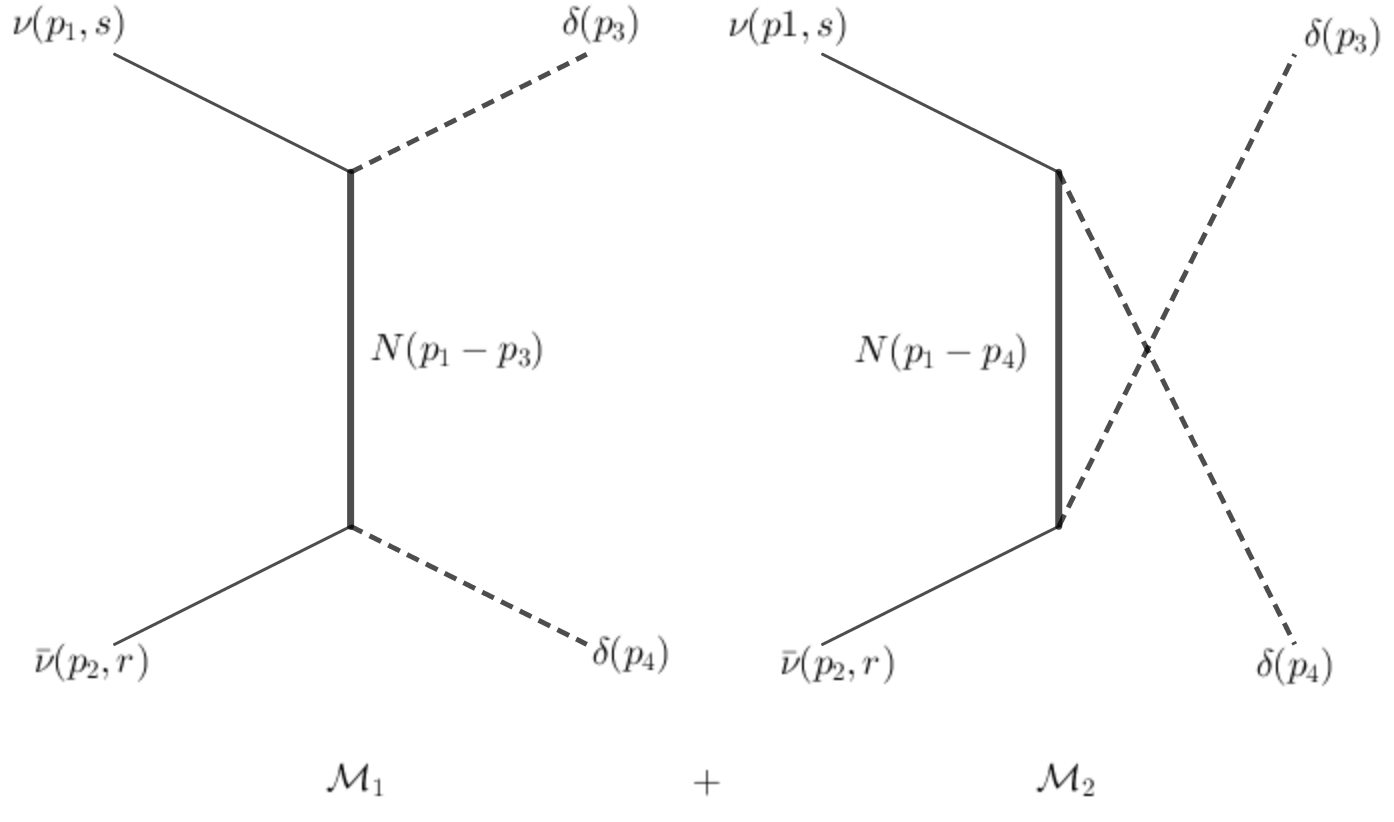}
\caption{The two diagrams contributing to $\nu \bar{\nu} \rightarrow \delta \delta$ scattering amplitude in the case that $\delta$ is a real scalar.}
\label{fig:feyn}
\end{figure}

We are interested in the processes of the form;
\begin{align*}
\nu \,\, \nu &\longrightarrow \delta \,\, \delta\\
\nu \,\, \overline{\nu} &\longrightarrow \delta \,\, \delta
\end{align*}
If the neutrinos are Majorana, both of these processes can occur. There are two diagrams contributing to the process, as shown in Figure \ref{fig:feyn}. The cross-section is given by;
\begin{multline}
\label{eq:sigma}
\sigma(s) = \frac{g^4 m_N^2}{32 \pi}\left[\sqrt{\frac{\frac{1}{4}s - m_\delta^2}{s}}\frac{2}{\left(m_\delta^2 - m_N^2 - \frac{1}{2}s\right)^2 - s\left(\frac{1}{4}s - m_\delta^2\right)} \right. \\ \left. + \frac{1}{s\left(m_\delta^2 - m_N^2 - \frac{1}{2}s\right)}\log\left(\frac{m_\delta^2 - m_N^2 - \frac{1}{2}s + \sqrt{s\left(\frac{1}{4}s - m_\delta^2\right)}}{m_\delta^2 - m_N^2 - \frac{1}{2}s - \sqrt{s\left(\frac{1}{4}s - m_\delta^2\right)}}\right)\right]
\end{multline}

\section{IceCube, IceCube-170922A, and TXS 0506+056}\label{sec:icecube}

The IceCube neutrino observatory \cite{IceCube} is located at the South Pole, consisting of one cubic kilometre of Antarctic ice. The goal of the project is to detect very high energy neutrinos from astrophysical sources. In this regard it is sensitive to energies in the range $300\,\text{GeV}$ to $1\,\text{EeV}$. The important specifics of the design are that it is a Cherenkov detector. When a muon neutrino interacts with the ice via charged current interactions, a muon is produced which emits Cherenkov radiation as it propagates through the medium. The energy of this muon, $\hat{E}_\mu$ is measured along with the propagation path. This can then be used to reconstruct;
\begin{itemize}
  \item The neutrino energy $E_\nu$, from the muon energy prior, $\hat{E}_\mu$ via for example Figure S5 in \cite{IceCube2018}. As a reference, in \cite{Kelly}, an estimated fit is given which can be used in simple practical cases;
  \begin{equation}
    \left(\frac{E_\nu}{\textrm{TeV}}\right) = 1.92\left(\frac{\hat{E}_\mu}{\textrm{TeV}}\right)^{1.14}
  \end{equation}
  \item The declination and right ascension of the original neutrino path.\footnote{One can access this data \href{https://icecube.wisc.edu/science/data/access}{\textit{here}}.} This is vital in matching up events with possible sources, and has lead to the deduction that the neutrino events we will be interested were most likely sourced from a blazar, TXS $0506$+$056$, at a redshift of $z \simeq 0.3365$.
\end{itemize}

\noindent The IceCube experiment is searching for high energy astrophysical neutrinos. Only very particular types of astrophysical object can produce such energetic particles. Of particular interest to us are objects known as \textit{blazars}. These are \textit{Active Galactic Nuclei (AGN)}, which consist of a supermassive black hole that converts the gravitational and rotational energy of accreting matter into highly relativistic jets \cite{Ackermann2018} pointing in our direction. In particular, IceCube believes to have detected high energy neutrinos from a known $\gamma$-ray source TXS $0506$+$056$. The key fact regarding this blazar is that it is at a redshift \cite{Kelly};
\begin{equation}
  z_{\mathrm{TXS}} \simeq 0.3365 \Rightarrow d_{\mathrm{TXS}} \simeq 1.3 \, \textrm{Gpc}
\end{equation}
\noindent where $d_{\mathrm{TXS}}$ is the comoving distance to the blazar.\footnote{$1\,\textrm{Gpc} \simeq 3.08567 \times 10^{27} \, \textrm{cm}$} This therefore presents a very interesting regime in which to test fundamental physics: we have $\textrm{TeV}$ - $\textrm{PeV}$ energy neutrinos propagating across gigaparsecs of distance. Along the way, we can therefore consider interactions with dark matter, the cosmic neutrino background etc. As we shall discuss further below, it also presents a prime example of the utility of \textit{multimessenger astronomy} \cite{Kelly, Ackermann2018} where neutrino events are calibrated with other $\gamma$ and X-ray experiments such as Fermi-LAT, H.E.S.S., and the Swift XRT.

To understand the relationship between the flaring $\gamma$-ray source TXS $0506$+$056$ and IceCube, one should note that on $22$ Septemeber $2017$, a neutrino with an energy of $\sim 290\,\textrm{TeV}$ was observed at IceCube (IceCube-170922A). This prompted a couple of responses;
\begin{itemize}
  \item \textit{Multimessenger Approach:} Immediately after the event, multiple collaborations began to establish the coincidence of the neutrino alert with the flaring state of TXS $0506$+$056$. Broadly this relies on correlating $\gamma$-ray and X-ray measurements of known catalogs of astrophysical objects with the angular position of the reconstructed neutrino path. For a more detailed account of this, see \cite{Ackermann2018}, however the salient point is that the chance coincidence is currently ruled out at $3-3.5\sigma$.
  \item \textit{Historical Approach:} It also provoked a search into data taken in $2014-15$ where it was found that $13 \pm 5$ excess events \cite{Kelly} appear to \textit{also} be coincident with TXS $0506$+$056$ in its flaring state. It thus appears that this blazar is a source of high energy astrophsyical neutrinos.
\end{itemize}
We will be interested in the high energy event to place our own constraints on the model under consideration. The careful analysis in \cite{Ackermann2018} provides a most probable energy of $290 \, \textrm{TeV}$ with a $90\%$ confidence level lower bound of $183 \, \textrm{TeV}$. More information on the analysis of the coincidence of IceCube-170922A with the flaring of TXS $0506$+$056$ as well as the historical data can be found in \cite{Ackermann2018, IceCube2018, Kelly, Padovani2018}. Furthermore, the viability of blazars as neutrino sources is discussed in \cite{Hooper2018, IceCube2018}.

\section{The Neutrino Luminosity from TXS 0506+056}\label{sec:nuluminosity}

We intend to use the mean free path, which we can define via;
\begin{equation}
  \ell^{-1} = \sum_{i}{n_{X_i} \sigma(\nu X_i \rightarrow Y_i)}
\end{equation}
to place bounds on the model. In order to use the observation of the $290 \, \textrm{TeV}$ neutrino as a constraint on fundamental interactions within the dark and neutrino sectors, we first must discuss the validity of comparing the mean free path to the comoving distance from the blazar. If the neutrino luminosity associated to the blazar is of a magnitude that saturates the maximum bound, we may deduce that the mean free path of the neutrinos is likely to be \textit{larger} than the comoving distance. To do so, we extract the 7.5 year upper bound on the neutrino luminosity as given in Figure 4 of \cite{Ackermann2018}. Referring to $1.7$ in \cite{dermer2009high}, the \textit{luminosity} radiated by a source between energies $\epsilon_1$ and $\epsilon_2$ is then simply;
\begin{equation}
  L[\epsilon_1, \epsilon_2] = 4\pi d_L^2 \int_{\log \epsilon_1}^{\log \epsilon_2}{\upd{\log\epsilon} \nu F_\nu }
\end{equation}
where $d_L$ is the photometric distance to the source. This upper bound is shown in Table \ref{tab:luminosity} along with the upper bound provided by the HAWC experiment on the photon flux. It should be emphasised that the main comparison point here is the neutrino luminosity, not the photon luminosity which is given to provide support for a given production mechanism. The value for the photons does not affect the conclusions regarding the bounds on the neutrino couplings. The inclusion of the photon flux will be discussed further in the next section. To proceed, we compare the neutrino luminosity with upper bounds derived in blazar modelling scenarios, in particular those given in \cite{Padovani2019}. Other references which make a detailed record of the relevant production mechanisms inside the jet environements of the blazar include \cite{Gao2018nat, Rodrigues2018, Padovani2018, Keivani2018}. 
\begin{table}[t]
\centering
\begin{tabular}{p{3.0cm}p{3cm}p{4cm}p{3.5cm}p{2.0cm}}
\toprule \textbf{Source} & \textbf{Energy Range} & \hfill \textbf{Luminosity} $\text{erg}\,\text{s}^{-1}$ & \textbf{Reference}\\
\midrule
\textit{Neutrino Source} & $186 \, \textrm{TeV}$ - $7.9 \, \textrm{PeV}$ & \hfill $\lesssim 10^{46}$ & \cite{Ackermann2018} \\
\textit{$\gamma$ Source (HAWC)} & $0.8 \, \textrm{TeV}$ - $74.0 \, \textrm{TeV}$  & \hfill $\lesssim 4.1 \times 10^{45}$ & \cite{Ackermann2018, HAWC} \\
\bottomrule
\end{tabular}
\caption{Numerical Results for the luminosity in the given energy range for the $\gamma$-ray and neutrino components of the blazar flux. Note that in the neutrino case, we are considering the 7.5 year exposure presented in \cite{Ackermann2018}}
\label{tab:luminosity}
\end{table}
We make the following observations. It might appear that there is a conflict between the upper bound of $\sim 10^{45} \,\textrm{erg}\,\textrm{s}^{-1}$ in \cite{Padovani2019} and that presented in Table \ref{tab:luminosity}. There are two reasons this is not the case;
\begin{enumerate}
    \item The value quoted in Table \ref{tab:luminosity} is an upper bound on the measurement due to the fact that an ensemble of distant sources may lead to a neutrino observation even if the expectation value for one source is very small \cite{Padovani2019}.
    \item The value quoted in \cite{Padovani2019} assumes that there is only one emmiting region within the blazar jet. Other studies such as \cite{Murase2014} find luminosities that saturate this upper bound due to multiple emitting regions.
\end{enumerate}
With this in mind, the data is consistent with the different theoretical predictions for the expected neutrino luminosity. Up to the uncertainty in this modelling, it appears that the measured luminosity is indeed close to saturating this bound. We then make a key deduction of this work, that therefore the mean free path of the 290 TeV neutrino is likely to be greater than the distance to the blazar. This will be the definition of our bound.

\subsection{Discussion regarding the consistency of the neutrino and photon flux}

In this subsection we would like to discuss the neutrino flux we use above and see how it compares to the observed photon flux.  The discussion here is a simple sanity check, this subsection therefore on its own contains no results which have any impact on the bound we obtain later. In particular, we are not using the photon flux to derive a bound, we just aim to discuss the discrepancy between the two fluxes.

Let us recall how neutrinos and photons are thought to be generated in the relativistic jets of active galaxies. In the hadronic scenario, highly boosted protons interact with photons in the jet from e.g. electron synchotron radiation. This leads to the production of neutral and charged pions via resonances (for example $p\gamma \rightarrow \Delta^+ \rightarrow p\pi^0$) or direct production (for example $p\gamma \rightarrow n\pi^+$). These highly relativistic pions then decay via $\pi^0 \rightarrow \gamma \gamma$ and $\pi^+ \rightarrow \ell^+ \nu_\ell$ where $\ell$ is a lepton \cite{Mucke:1998mk, Szabo:1994qx}. This leads to a production of neutrinos and photons with $F_\nu \sim F_\gamma$ within the jet \cite{Keivani2018}. We might expect that the detection of high energy neutrinos should thus be accompanied by the EM emission of pionic gamma-rays. If this were the case, we would indeed expect the luminosities of the neutrinos and the photons to be comparable, $F_\nu \sim F_\gamma$, as was assumed in \cite{Kelly}.  Unlike the HAWC constraint, the Fermi-LAT data consisted of an actual measurement, and those authors used this assumption to make a direct prediction for the neutrino flux.

\begin{figure}[t]
 \centering
 \includegraphics[width=.5\textwidth]{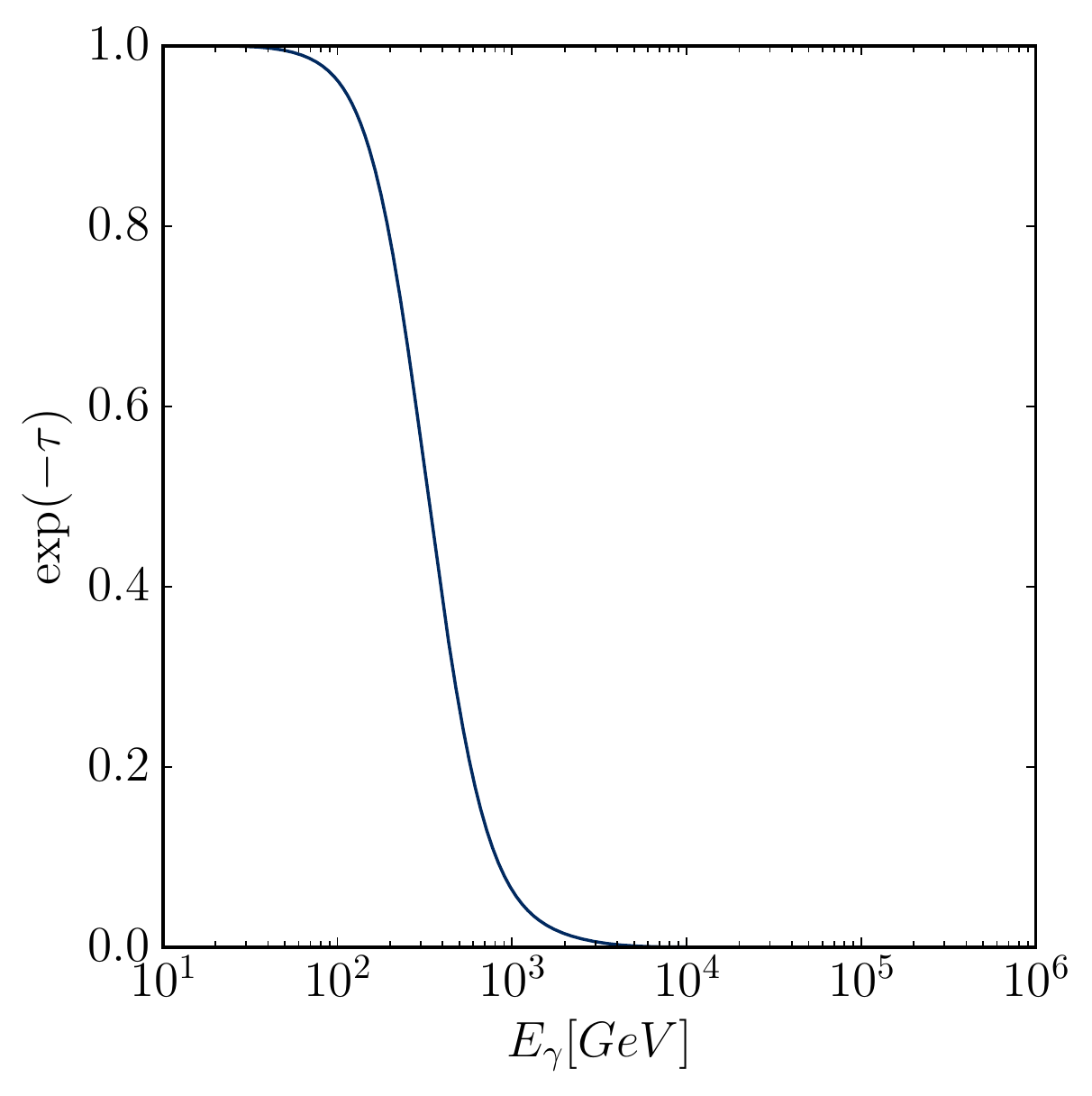}
 \caption{The probability, $\exp(-\tau)$, where $\tau$ is the optical depth, of a photon produced in the blazar jet reaching the Earth due to interactions with the EBL. We see that at Fermi-LAT energies, $\mO(290) \, \mathrm{GeV}$, this probability is close to 1, whilst at the higher, HAWC energies, $\mO(1 - 100) \, \mathrm{TeV}$, there is significant attenuation of the flux due to scatterings $\gamma\gamma \rightarrow e^{+}e^{-}$. These calculations are based on \cite{DeLavallaz:2011ju}.}
 \label{fig:ebl}
\end{figure}

The assumption that $F_\nu \sim F_\gamma$ is however conservative --- since the neutrinos are weakly interacting, they escape the jet without attenuation to the flux. On the other hand, the photons produced by neutral pion decays, may \textit{not} be observed due to electromagnetic processes which may occur in the jet or attenuation during propagation across on the Universe. In the latter case this is due to $\gamma\gamma \rightarrow e^+ e^-$ attentuation on the Extragalactic Background Light (EBL) \cite{Finke:2009xi}.  In Figure \ref{fig:ebl} we use code developed by one of the authors for a previous project \cite{DeLavallaz:2011ju} to show the attentuation due to pair production on the EBL for high energy photons from a blazar at redshift $z=0.34$ is not very important at the Fermi-LAT energies considered in \cite{Kelly} ($<$ 290 GeV) but really cuts off the photon flux at the HAWC energies relevant here (0.8 TeV --- 74 TeV). Because of this, the HAWC data acting as an upper bound is in no conflict with the jet physics.

The HAWC data in Table \ref{tab:luminosity} shows that the photon flux at this energy is less than the neutrino flux we have assumed.  Given the fact it is much easier for photons to be attenuated and to lose energy than neutrinos, we assume this is in fact what has happened and note that we have assumed the lower of the two possible estimates of the neutrino flux based on the observed event.

\section{Assumptions in our Approach}\label{sec:assumptions}

After setting up the context for the model, the experiment, and the broad idea of the analysis, we now fill in some of the details regarding the assumptions in our approach. In this section we will consider (i) the sub-dominance of other particle processes, (ii) redshift effects, (iii) neutrino sector technicalities such as mass hierarchies, and, (iv) the non-degeneracy of the scalar mass eigenstates.

\subsection{Why do we only have to worry about one process?}\label{sec:oneprocess}

In Section \ref{sec:crosssection} where we detailed how to do the calculation for $\nu\nu \rightarrow \delta\delta$, we neglected to calculate the cross section for other processes within the model that may also lead to a neutrino interaction. These additional processes are as follows;
\begin{enumerate}
  \item \textit{$\nu\nu \rightarrow NN$:} One can construct scenarios in parameter space where this dominates. However the centre of mass energy: $E_{\mathrm{com}} = \sqrt{2 E_\nu m_\nu}$ where $m_\nu$ is the mass of the cosmic neutrino background neutrino, is close to the mass of the lightest scalar $\delta$. By construction if the scalar is the dark matter candidate, then the mass of $N$, $m_N$, must be larger. Thus, even in scenarios where the centre of mass energy is large enough to produce two $N$ particles, the cross section is likely to lie at the front tail of the distribution and so be subdominant.
  \item $\nu N \rightarrow \nu N$: There is an $s$-channel and a $t$-channel diagram for this process. Independent of this however, we have assumed that $N$ is not the dark matter candidate. As such, we expect the relic density to be very low in comparison to all other particles. The vertex structure ensures that we expect the cross section to be of a similar order of magnitude to the $\nu\nu \rightarrow \delta\delta$ case. Hence, the contribution to the mean free path is negligible.
  \item $\nu\delta \rightarrow \nu\delta$: It is not immediately clear as to whether this will be negligible. Firstly, there is a $t$-channel process that will have a similar algebraic cross section as previously calculated in $\nu\nu \rightarrow \delta\delta$. There is also an $s$-channel process, whose cross-section we obtain from \cite{Franarin2018}. Secondly, we must check the contribution to the mean free path in two regimes;
  \begin{itemize}
    \item On a cosmological scale where $n_\delta$ is given by the relic density of dark matter,
    \item On a galactic scale, where the density is much higher within the dark matter halo.
  \end{itemize}
\end{enumerate}
Note that neglecting these processes is of course a simplifying assumption about the nature of the cross-sections, but they do not affect the interpretation of the results. This is because including any of the additional contributions above can only \textit{improve} the bounds; for a given $\ell$, introducing a new process increases the effective cross section, and reduces the mean free path leading to tighter constraints.
\subsection{Contribution from $\nu\delta \rightarrow \nu\delta$}
As mentioned above, we must check whether the $\nu\delta \rightarrow \nu\delta$ process contributes significantly to the mean free path of the blazar neutrino. In what follows, we will find that it does not contribute significantly. This is due to the fact that, even within the galactic halo, the cross-section is too small to generate a significant contribution.
\subsubsection{$t$-Channel Cross Section}\label{sec:tchannel}

The relationship;
\begin{equation}
    u = m_\delta^2 - \frac{1}{2}s - \sqrt{s\left(\frac{1}{4}s - m_\delta^2\right)}\cos\theta
\end{equation}
along with the observation that for $\nu\delta \rightarrow \nu\delta$ with $m_\delta \sim \mO(\mathrm{MeV})$ and $E_\nu \sim \mO(TeV)$, it follows that $s \sim \mO(GeV) \gg m_\delta$. This then implies in this energy regime, $u \simeq s$. By crossing symmetry, we then deduce that the dependence of $\sigma_t(\nu\delta \rightarrow \nu\delta)$ on the centre of mass energy is just given by $\sigma(s)$ as in \eqref{eq:sigma}. To compare $\sigma(\nu\nu \rightarrow \delta \delta)$ and $\sigma_t(\nu\delta \rightarrow \nu\delta)$ we need only to evaluate $\sigma(s)$ at the different centre of mass energies, $s = 2E_\nu m_\nu$ and $s = 2 E_\nu m_\delta$. For an explicit comparison, we put in the values $g_e = 3 \times 10^{-3}$, $g_\mu = 10^{-2}$, $g_\tau = 3\times 10^{-1}$, $m_\delta = 0.5 \, \textrm{MeV}$, $m_\nu = 0.15 \, \textrm{eV}$, $m_N = 5\,\textrm{MeV}$. We find;
\begin{equation}
  \sigma(\nu\nu \rightarrow \delta\delta) \simeq 4.7\times 10^{-10} \, \textrm{MeV}^{-2}, \quad \sigma_t(\nu\delta \rightarrow \nu\delta) \simeq 6.2\times 10^{-16}\, \textrm{MeV}^{-2}
\end{equation}
So we find there is a difference of five to six orders of magnitude. After computing the number density of the dark matter in the galactic and cosmological cases, we will use this to deduce that the $t$-channel does not contribute.

\subsubsection{$s$-Channel Cross Section}\label{sec:schannel}
We obtain an analytic expression for the $s$-channel cross-section from \cite{Franarin2018};
\begin{equation}
  \sigma_s(\nu_\mu\delta \rightarrow \nu_\ell \delta) = \frac{g_\mu^2 g_\ell^2}{16\pi}\frac{(m_N^2 - m_\delta^2)^2}{m_N^2 + m_\delta^2} \frac{1}{(s - m_N^2)^2 + \Gamma_N^2 m_N^2}
\end{equation}
where $\Gamma_N$ is the width of $N$, it is given by;
\begin{equation}\label{eq:sigma_s}
  \Gamma_N = \sum_{\ell}{\frac{g_\ell^2}{16\pi} \frac{(m_N^2 - m_\delta^2)^2}{m_N^3}}
\end{equation}
In this case, we simply do an order of magnitude estimate with;
\begin{equation}
  m_N = \mO(\textrm{MeV}), \quad m_\delta = \mO(\textrm{MeV}), \quad s - m_N^2 = \mO(\textrm{GeV}^2), \quad g = \mO(10^{-2})
\end{equation}
We note that this implies that $(s - m_N^2)^2 \gg \Gamma_N^2 m_N^2$. Putting these into \eqref{eq:sigma_s}, we find;
\begin{equation}
  \sigma_s(\nu\delta \rightarrow \nu\delta) \simeq \mO(10^{-21}\,\textrm{MeV}^{-2})
\end{equation}
As such we deduce that the contribution is certainly negligible at this energy, since even if the number density of $\delta$ was high enough, the $t$-channel process will dominate by $4$ or $5$ orders of magnitude. 

\subsubsection{The interference term}\label{sec:interference}
To argue that the interference term between the $s$ and the $t$ channel amplitudes, which we denote $\mM_s$ and $\mM_t$ respectively, also leads to a negligible contribution to the cross section, we note that by the triangle inequality;
\begin{equation}
    \Abs{\mM_s + \mM_t} \leq \Abs{\mM_s} + \Abs{\mM_t} \Rightarrow \Abs{\mM_s + \mM_t}^2 \leq \Abs{\mM_s}^2 + \Abs{\mM_t}^2 + 2\Abs{\mM_s}\Abs{\mM_t}
\end{equation}
Furthermore, integrals of these quantities which ultimately lead to the total cross section will satisfy equivalent relations due to the positive definite nature of the integrands. Finally then, if we denote the amplitude for $\nu\nu \rightarrow \delta\delta$ as $\mM$, the calculations in the previous two sections indicate that $\Abs{\mM_s}^2 \sim \mO(10^{-10})\Abs{\mM}^2$ and $\Abs{\mM_t}^2 \sim \mO(10^{-6})\Abs{\mM}^2$. Hence, by the triangle inequality, we deduce that $\Abs{\mM_s}\Abs{\mM_t} \sim \mO(10^{-8})\Abs{\mM}^2$ and therefore leads to a neglible contribution to the total squared amplitude, $\Abs{\mM_s + \mM_t}^2 \sim \Abs{\mM_t}^2 \sim \mO(10^{-6})\Abs{\mM}^2$. Hence we deduce that the total cross section for $\nu\delta \rightarrow \nu\delta$, where $\nu$ in the initial state has energy $\mO(\mathrm{TeV})$ satisfies $\sigma(\nu\delta \rightarrow \nu\delta) \sim 10^{-6} \cdot \sigma(\nu\nu\rightarrow \delta\delta)$. It therefore remains to check the number density of dark matter in the cosmological and galactic cases and compute the mean free path using the $t$-channel cross-section.

\subsubsection{Number Density on Cosmological Scales}
The dark matter density on a cosmological scale, at a redshift $z$, is given by \cite{Farzan2014};
\begin{equation}
  n(z) = \frac{\Omega_{\textrm{DM},0}\rho_c}{m_{\textrm{DM}}}(1 + z)^3 \simeq 1.26 \times 10^{-3} (1 + z)^3 \left(\frac{\textrm{MeV}}{m_{\textrm{DM}}}\right) \, \textrm{cm}^{-3}
\end{equation}
where we have used $\Omega_{\textrm{DM},0} \simeq 0.265$ and $\rho_c = 3H_0^2/8G \simeq 4.77\, \textrm{keV}\,\textrm{cm}^{-3}$. This is $5$ orders of magnitude below the cosmic neutrino background number density. So in order to be relevant, $\sigma(\nu\delta \rightarrow \nu\delta)$ would have to be approximately $10^5$ times larger than $\sigma(\nu\nu \rightarrow \delta\delta)$, evaluated at the neutrino energy. From Sections \ref{sec:tchannel}, \ref{sec:schannel}, and \ref{sec:interference}, we see this is not the case, and indeed the cross-section is significantly smaller than the $\nu\nu \rightarrow \delta\delta$ cross-section.  We can therefore neglect this cross section as the neutrino travels to the Milky Way.

\subsubsection{Number Density on Galactic Scales}
After deducing that the contribution to the mean free path from interaction with dark matter is negligible on cosmological scales, we just have to check whether the galactic overdesnity could lead to a significant contribution. As in \cite{Franarin2018}, we use the Einasto profile with $\alpha = 0.15$ and $R_0 = 20\,\textrm{kpc}$ to model the dark matter energy density;
\begin{equation}
  \rho_{\textrm{DM}}(r) = 7.2 \times 10^{-2}\,\textrm{GeV} \, \textrm{cm}^{-3}\, \cdot \exp\left(-\frac{2}{\alpha}\left(\left(\frac{r}{R_0}\right)^\alpha - 1\right)\right)
\end{equation}
We can obtain the number density by dividing by the mass of the dark matter particle, $m_\delta \simeq \mO(10^{-3} \, \textrm{GeV})$. Now, note that this is maximal when $r = 0$. In order to put an upper bound on the contribution to the optical depth, we assume that the whole halo has this maximal number density. With $m_\delta = 1\, \textrm{MeV}$;
\begin{equation}
  n_{\textrm{DM}}^{\textrm{max}} \simeq \frac{\rho_{\textrm{DM}}(r = 0)}{m_\delta} \simeq 4.4 \times 10^{7} \, \textrm{cm}^{-3} \simeq 10^{5} n^0_{\nu}
\end{equation}
Now we are in a position to see why this does not contribute to the suppression of the neutino flux from the blazar. Whilst the combination of $n^{\textrm{max}}_{\textrm{DM}} \sigma_t(\nu\delta\rightarrow\nu\delta)$ is now of the same order of magnitude as $n_\nu \sigma(\nu\nu\rightarrow \delta\delta)$, the relevant consideration is the probability that such an interaction ($\nu\delta \rightarrow \nu\delta$) occurs. This is dependent on the ratio between the length scale at which such a high number density is observed (i.e. the galactic radius), and the mean free path. Here, the mean free path, $\ell$, is $\mO(\textrm{Gpc})$, so the probability of survival is $\sim \exp(-d_{g}/\ell)$ where $d_g \simeq 1\,\textrm{kpc}$ is the galactic radius. We see that this is approximately unity. Finally, note that in the case where the mean free path is $\mO(\textrm{kpc})$, we would not expect the neutrino to reach anywhere close to the galaxy, so this would be inconsequential also. Hence we deduce that both in the cosmological and galactic settings, the contribution from $\nu\delta \rightarrow \nu\delta$ is negligible in comparison to the dominant $t$-channel process $\nu\nu \rightarrow \delta\delta$.

To end this section, we emphasise that whilst we have neglected the contribution from these other processes, including any/all of them can only improve the bounds as the mean free path will decrease with new interactions. As such, it is only for the sake of simplicity that we make the assumptions, \textit{not} at the cost of the validity of the bounds.

\subsection{Mass Splitting in the Complex Case}\label{sec:complexsplit}

We assume that each of the mass eigenstates is equally abundant $\nu_1, \bar{\nu}_1, \nu_2, \ldots$, with a number density given by;
\begin{equation}
  n_{\nu_i} = \frac{1}{6} n_\nu = \frac{1}{6} \cdot 340 \, \textrm{cm}^{-3}
\end{equation}
Now, the contribution from each mass eigenstate to the inverse mean free path $\ell^{-1}$ is given by $n_{\nu_i} \sigma(\nu_\mu X \rightarrow Y )$. Importantly we argued in the last section that we thus need only consider $\sigma(\nu_\mu \nu \rightarrow \delta \delta)$. Now, in the real case, there is nothing more to say as there is only one scalar mass eigenstate. In the complex case however, we must consider the following. The theory we are considering is effective up to some scale $\Lambda$. It therefore does not have to explicitly respect any of the symmetries that might apply in the UV. Indeed all we assume is that the new dark sector particles are odd under a $\mathbb{Z}_2$ symmetry, to ensure there is a stable candidate. As such, writing $\delta = \tfrac{1}{\sqrt{2}}(\delta_1 + i \delta_2)$, the most general hermitian mass term can be written;
\begin{equation}
  V_m = M^2 \delta\dagg \delta - \frac{1}{2}(m^2 \delta \delta + \textrm{h.c.})
\end{equation}
This leads to a mass splitting between the mass eigenstates $\delta_{1,2}$ given by $\Delta m_{12}^2 = 2m^2$. Now, we consider the possible processes $\nu\nu \rightarrow \textrm{scalars}$. We have;
\begin{equation*}
\nu\nu \rightarrow \delta_1 \delta_1, \quad \nu\nu \rightarrow \delta_1 \delta_2, \quad \nu\nu \rightarrow \delta_2 \delta_2
\end{equation*}
From this we see that there are a couple of scenarios that might occur kinematically. We assume that the lightest scalar is $\delta_1$, and that the first process can happen. Then it may the case that either (i) only the first process can occur, (ii) only the first and second processes can occur, or, (iii) all the processes can occur. This is where we make our simplifying assumption, which unlike the first case, will not necessarily improve the bounds if put in at a later date. We assume that if the first occurs, then the next two may also occur. This is equivalent to saying that there is a small mass gap between the two eigenstates. To simplify the situation then we assume that $m$ is small compared to $M$, and therefore that we can approximate;
\begin{equation}
  \sigma(\nu\nu \rightarrow \delta_1 \delta_1) + \sigma(\nu\nu \rightarrow \delta_1 \delta_2) + \sigma(\nu\nu \rightarrow \delta_2 \delta_2) \simeq 3 \sigma(\nu\nu \rightarrow \delta_1 \delta_1)
\end{equation}

\subsection{Redshift Considerations}

During the cosmological propagation, both the number density of the cosmic neutrino background neutrinos, and the energy of the blazar neutrino will be affected by redshift. Let the values now be denoted $n_\nu^0$ and $E_\nu^0$ respectively, then at a redshift $z$;
\begin{equation}
  n_\nu(z) = n_\nu^0 (1 + z)^3, \quad E_\nu(z) = (1 + z)E_\nu^0
\end{equation}
We now make the observation that the source of the $290 \, \textrm{TeV}$ neutrino is at a redshift $z = 0.3365$. In the case of the energy this means that the maximum possible multiplicative factor is $(1 + 0.3365)$, but this is within the confidence bounds on the energy measured at IceCube, so can be neglected. The redshift of the number density is not negligible however, although it only improves the bounds. We take the result from \cite{Farzan2014} that the optical depth is given by;
\begin{equation}
  \tau = c\int_{z = z_1}^{z = z_2}{\upd{z}\frac{\ud t}{\ud z}n(z)\sigma(z)}
\end{equation}
Now, the energy of the muon observed at IceCube had a 1$\sigma$ confidence interval of $23.7 \pm 2.8$ TeV \cite{IceCube2018}. This can be translated into an error on the energy of the incoming neutrino of an order 100 TeV. With this observation, we note that the energy of the 290 TeV neutrino at its source, i.e. before it is redshifted during the propagation, will lie within these bounds. Therefore, to simplify the analysis, we assume that $\sigma(z) = \sigma(E_\nu(z))$ does not depend on the redshift, $z$. With this assumption in mind, the expression above reduces to; 
\begin{equation}
  \tau = c n_\nu^0 \sigma(E_\nu^0) \int_{z = z_1}^{z = z_2}{\upd{z}(1 + z)^3 \frac{\ud t}{\ud z}}
\end{equation}
We can relate $\ud t/\ud z$ to the Hubble rate via;
\begin{equation}
  \frac{\ud t}{\ud z} = -\frac{1}{(1 + z)H(z)}
\end{equation}
where;
\begin{equation}
  H(z) = H_0 \sqrt{\Omega_\Lambda + \Omega_{m,0}(1 + z)^3}
\end{equation}
We will take the values $\Omega_\Lambda \simeq 0.65$, $\Omega_{m,0} \simeq 0.315$, $H_0 \simeq 6.73 \times 10^{4}\, \textrm{km}\,\textrm{s}^{-1}\textrm{Gpc}^{-1}$ \cite{Planck}, $c = 3\times 10^{5}\,\textrm{km s}^{-1}$. Letting $\ell_0^{-1} := n_\nu^0 \sigma(E_\nu^0)$ to find;
\begin{equation}
  \tau = \left(\frac{\ell_0}{\text{Gpc}}\right)^{-1} \left(\frac{c \, / \, \text{km s}^{-1}}{H_0 \, /\, \text{km s}^{-1}\text{Gpc}^{-1}}\right) \cdot \int_{z = z_1}^{z = z_2}{\upd{z}\frac{(1 + z)^2}{\sqrt{\Omega_\Lambda + \Omega_{m,0}(1 + z)^3}}} \simeq 1.90 \left(\frac{\ell_0}{\textrm{Gpc}}\right)^{-1}
\end{equation}

\subsection{Including Neutrino Mass Hierarchies}

The last technicality to introduce into the computation of the bounds are the facts that;
\begin{enumerate}
  \item The neutrino \textit{mass} eigenstates and \textit{flavour} eigenstates are not the same
  \item The neutrino masses are unknown, and indeed have two possible orderings (for the mass eigenstates); the \textit{normal hierarchy} and the \textit{inverted hierarchy}.
\end{enumerate}
Within our calculation we aim to present the situation for both of these cases.

\subsubsection{Mass Eigenstates and the PMNS Matrix}

Within the Standard Model, we expect neutrinos to be massless. Experiments illustrating phenomena such as neutrino oscillations contradict this fact and we now believe they do indeed have a small mass. This complicates matters however for the reason mentioned above. The flavour eigenstates and the mass eigenstates are no longer the same in this case. Instead they are related by the \textit{PMNS} matrix\footnote{Pontecorvo-Maki-Nakagawa-Sakata}. This encodes a unitary transformation between the flavour basis and the mass basis:
\begin{equation}
\nu_\ell := \begin{pmatrix} \nu_e \\ \nu_\mu \\ \nu_\tau \end{pmatrix} = \thrbythr{U_{e1} & U_{e2} & U_{e3}}{U_{\mu1} & U_{\mu2} & U_{\mu3}}{U_{\tau1} & U_{\tau2} & U_{\tau3}} := U \nu_i
\end{equation}

\subsubsection{The Mass Hierarchy}

A key fact in this discussion is that ultimately we do not know the absolute values, nor the ordering of the mass eigenstates. There are two common alternatives, which are illustrated in Figure 2 in \cite{King};
\begin{enumerate}
  \item \textit{Normal Ordering:} In the normal hierarchy, $\nu_3$ is the most massive state, whilst $\nu_1$ and $\nu_2$ are lighter.
  \item \textit{Inverted Ordering:} On the other hand, in the inverted case, $\nu_3$ is the lightest, whilst $\nu_1$ and $\nu_2$ are heavier.
\end{enumerate}

\subsubsection{Constraints on the Masses}

We can go slightly further, whilst we do not know the precise masses of the neutrinos we have (i) a bound on the total sum of the masses that comes from Cosmology, and, (ii) values for the mass difference between the eigenstates. To be more precise;
\begin{itemize}
  \item Combining constraints from Cosmic Microwave Background (CMB) anisotropies, Baryon Acoustic Oscillations, Type 1A Supernovae, and, CMB lensing, we will use the constraint \cite{Couchot2017};
  \begin{equation}
    \sum{m_{\nu_i}} < 0.17 \, \textrm{eV}
  \end{equation}
  \item We also know the squared mass differences between some of the mass eigenstates \cite{Couchot2017};
  \begin{align}
    \Delta m_{12}^2 = m_2^2 - m_1^2 &= 7.37 \times 10^{-5}\,\textrm{eV}^2 \\
    \Delta m^2 = m_3^2 - \frac{1}{2}(m_1^2 + m_2^2) &= +2.50 \times 10^{-3} \, \textrm{eV}^2\, \textrm{(NH)} \\
    &= -2.46 \times 10^{-3} \, \textrm{eV}^2\, \textrm{(IH)}
  \end{align}
\end{itemize}
From the last of these constraints we see that fixing one of the masses automatically fixes the others. In our analysis we intend to vary one of the masses of the mass eigenstates and use the squared mass differences to compute the other masses, remaining within the bound set by the cosmological considerations. We will present the analysis in both the normal and inverted cases.

\subsection{The Coupling Constants}

There is one final consequence of the non-coincidence of the mass and flavour eigenstates. We are considering a coupling in the Lagrangian of the form;
\begin{equation}
  \mL_{\textrm{new}} = \sum_{\ell}{g_\ell \delta \bar{N}_R \nu_{\ell, L} + \textrm{h.c.}}
\end{equation}
where importantly, the $\nu_{\ell}$ are the flavour eigenstates. Furthermore, we quoted constraints on the couplings $g_\ell$ in this flavour basis e.g. $g_{\ell} < 10^{-3}$ in the case of real dark matter. Now consider expanding in the mass basis;
\begin{equation}
  \mL = \delta \bar{N}_R \sum_{\ell}{g_\ell \sum_{i}{U_{\ell i}\nu_{i, L}}} + \textrm{h.c.} := \sum_i{g_i \delta \bar{N}_R \nu_{i, L}}
\end{equation}
We have defined the couplings to the neutrino mass eigenstates;
\begin{equation}
  g_i := \sum_{\ell}{U_{\ell i}g_{\ell}}
\end{equation}
Now, importantly, these will inherit constraints from the constraints on the flavour basis couplings, and are just related by a linear transformation. This means that we can still parametrise our constraints in terms of the flavour couplings. The context of these comments is that the cosmic neutrino background consists of decoherent mass eigenstates. Therefore instead of considering flavour processes $\nu_\mu \nu_\ell, \nu_\mu \bar{\nu}_\ell \rightarrow \delta\delta$, we should instead consider $\nu_\mu \nu_i \rightarrow \delta \delta$. To do so we should use the $\set{g_i}$ couplings at the $\nu_i \delta N$ vertex, which we can compute as above. We also make use of the mass eigenstate masses as discussed above to compute the centre of mass energy in each of the different cases $i = 1,2,3$. Finally, we will assume that each of the mass eigenstates is equally abundant in the cosmic neutrino background so that we can take the number density of each species to be $n_\nu/6$ as noted previously.

\subsection{Neutrino Clustering}

This is the phenomenon relating to the gravitational clustering of neutrinos at late times once they become non-relativistic. This can increase their density inside gravitational wells such as the Milky Way today.  An important reference is \cite{Ringwald2004} which discusses the clustering of cosmic neutrino background neutrinos onto cold dark matter. In the context of this work, this would affect the number density $n_\nu(z)$ as the astrophysical neutrino passed through different dark matter distributions. In regions where there is more cold dark matter, \cite{Ringwald2004} suggests that we should also see more cosmic neutrino background neutrinos. A precision analysis of the propagation of the neutrinos from the blazar should take this into account.

This being said, \cite{Ringwald2004} only extends the analysis to the local group\footnote{The \textit{Greisen-Zatsepin-Kuzmin} zone}, across distances of $\textrm{Mpc}$. This is ultimately small scale structure in the context of $\textrm{Gpc}$ propagation. Figure 8 in \cite{Ringwald2004} illustrates the density contrast of the neutrinos on this scale. We see that density constrasts of $\mO(2)$ are realistic, so including this effect could strengthen the bounds. Even an increase of an order of magnitude within the local group would only change the optical depth at the percent level, so we neglect this effect in this work.

\section{Results}\label{sec:results}

We are now in a position to discuss the results. First we shall present the methodology applied, in particular what parameter space we explore, and which effects are included in the analysis. For a given set of parameters ($g_{ie}, \cdots, m_\delta, m_N$), the approach taken is as follows;
\begin{enumerate}
  \item Consider the scattering of a $290 \, \textrm{TeV}$ muon-neutrino off the mass eigenstates in the cosmic neutrino background.
  \item Compute the cross-section at the centre of mass energy \emph{for each} mass eigenstate. This allows us to include the different neutrino hierarchies.\footnote{In the complex case, we include an additional factor of $3$ as discussed in Section \ref{sec:complexsplit}}
  \item Compute the mean free path, including the redshifting effect and compare to the distance to the blazar.
  \item Reject parameters sets for which the mean free path is less than this distance.
\end{enumerate}

\subsection{Parameter Choices}

In terms of the parameters we choose, we are informed by the current leading observational bounds on (i) the total neutrino mass, and, (ii) the coupling constants in the effective theory. Before proceeding, we make it clear that we could not improve on the constraints $g_{\ell} \lesssim 10^{-3}$ in the case of real dark matter. The mean free path was always significantly larger than the distance to the blazar. All the results that follow are in the complex case, where for example, the coupling to the tau neutrino is far less constrained. Taking the constraints from \eqref{eq:constraints}, we choose the following for our computations;
\begin{itemize}
  \item The mean free path is most sensitive to the coupling, $g_{i\mu}$, as such we choose to vary this parameter in the range $g_{i\mu} \in [10^{-5}, 10^{-1}]$.
  \item We also vary the mass of the mediator $m_N \in [m_\delta, 10\,\textrm{MeV}]$.
  \item We fix the mass of the scalar $m_\delta$ for values $m_\delta = 0.1, 0.5, 1.0, 1.5 \, \textrm{MeV}$.
  \item The electron neutrino coupling is the most constrained, and indeed we find that the conclusions are insensitive to this parameter. As such, we set $g_{ie} = 0$, removing the dependence.
  \item In all the plots below we have take $g_{\tau} = 3\times 10^{-1}$, at the upper bound of the current constraints. We acknowledge that this may be slightly optimistic.
  \item Finally we consider values of the lightest neutrino (in both hierarchies) of $m_\nu = 0.01, 0.03, 0.04\,\textrm{eV}$. All of these values are within the Planck bound of $\sum_{\nu}{m_\nu} < 0.17 \, \textrm{eV}$. Although, since this bound appears to be getting tighter, the lower neutrino masses e.g. $m_\nu = 0.01 \Rightarrow \sum_{\nu}{m_\nu} \simeq 0.06 \, \textrm{eV}$ perhaps lead to stronger conclusions.
\end{itemize}

\subsection{Analysis}
A selection of results are shown in Figure \ref{fig:main} across a range of neutrino and scalar masses. A complete set of plots are shown in the appendix, in Figure \ref{fig:appendix}. Note that the shaded regions in each figure are those that are ruled out by the mean free path, relic abundance, and kaon decay constraints. For example, masses of the mediator above $10 \, \textrm{MeV}$ are ruled out by the thermal cross-section calculations.
\begin{figure}[t]
 \centering
 \includegraphics[width=.48\textwidth]{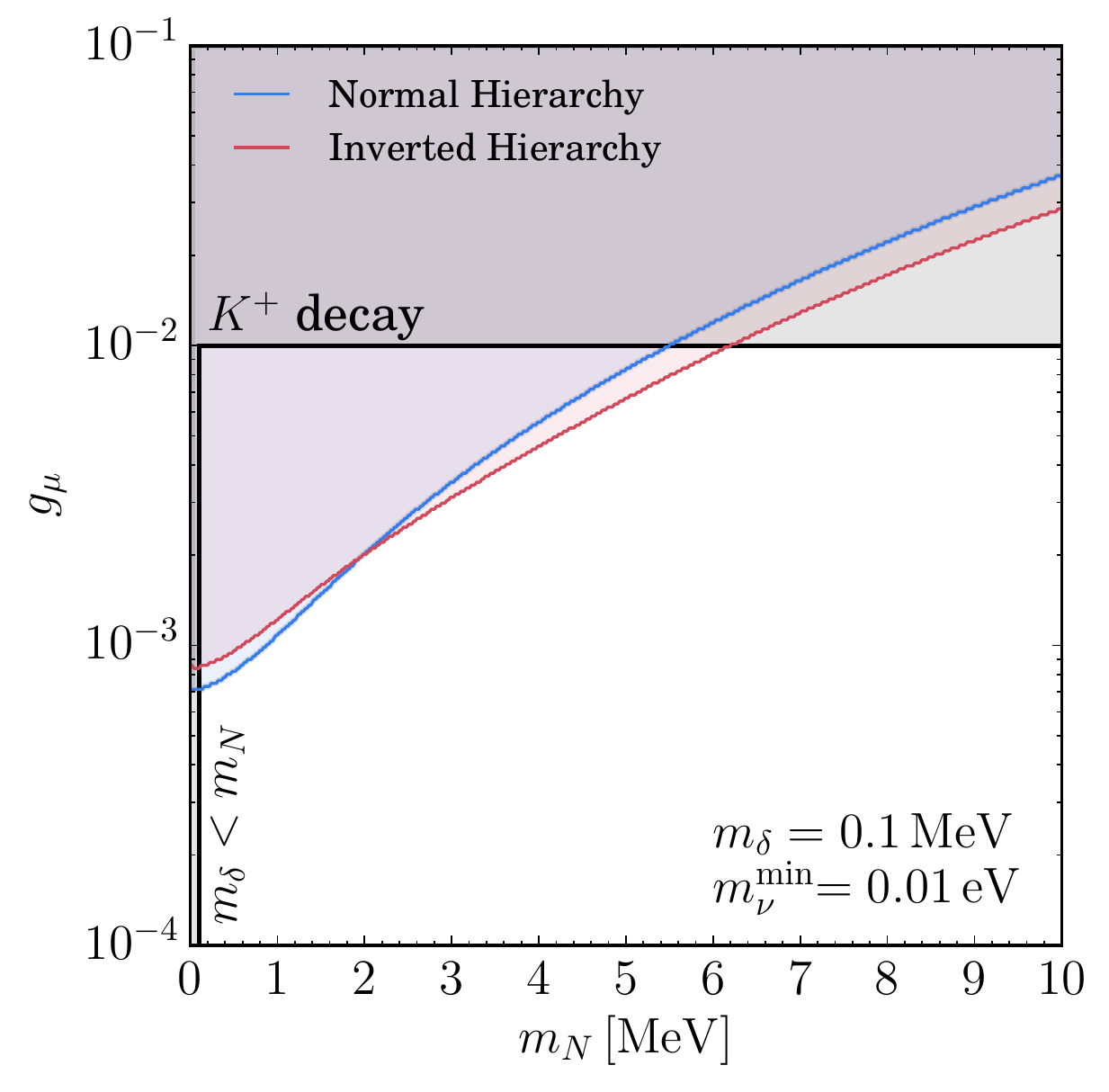}\quad
 \includegraphics[width=.48\textwidth]{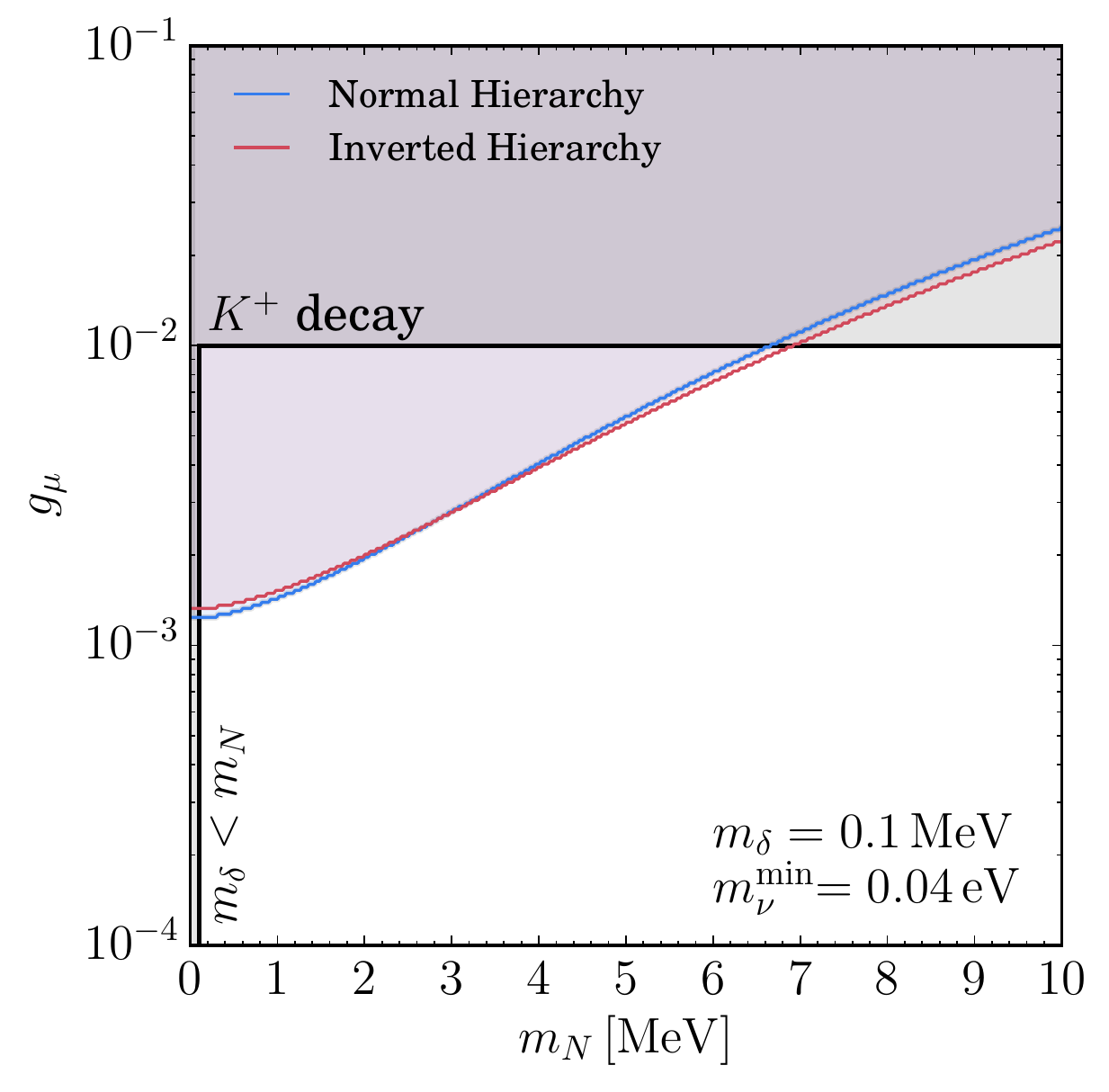}

 \medskip

 \includegraphics[width=.48\textwidth]{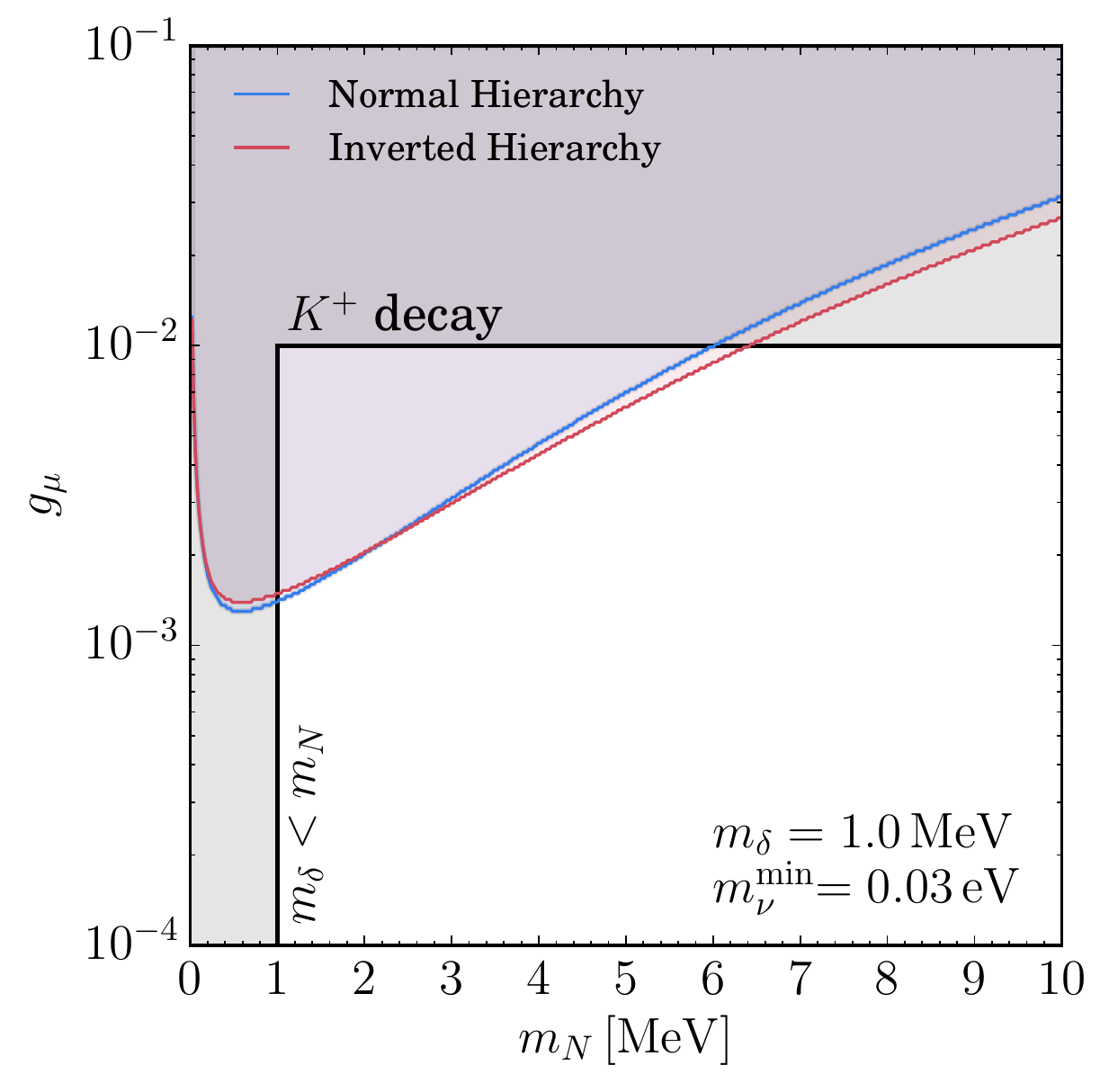}\quad
 \includegraphics[width=.48\textwidth]{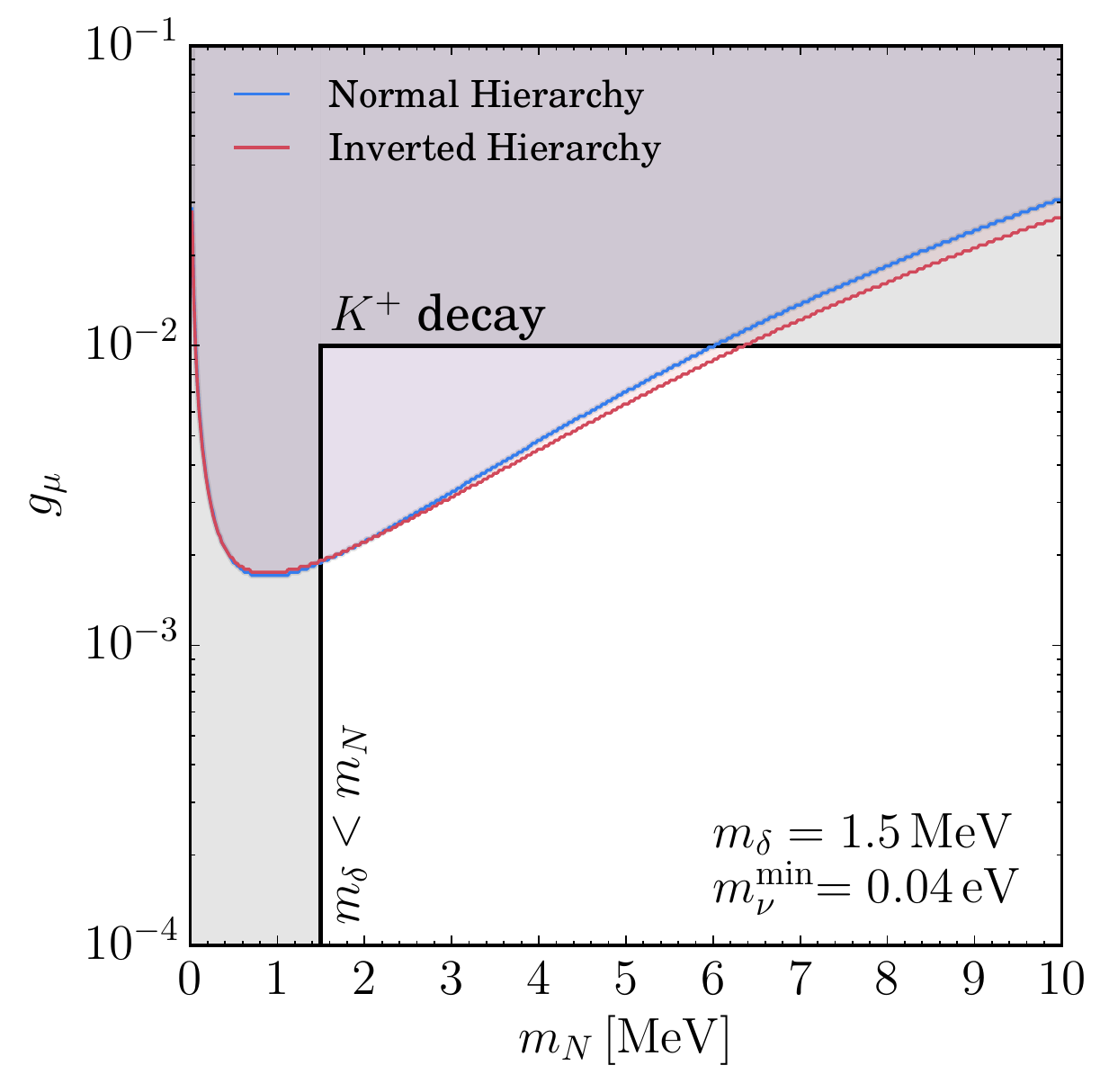}
 \caption{Selection of constraint plots for the case of complex scalar dark matter. In each plot, $g_e = 0$, $g_\tau = 3 \times 10^{-1}$, and the shaded regions are those that are ruled out by the mean free path constraints.}
 \label{fig:main}
\end{figure}
\noindent As we will discuss shortly, the constraints we obtain in this way are stronger than those from Kaon decays, but slightly weaker than the constraints from BBN and CMB, we shall discuss this in more detail very shortly.  We see from the plots that at low mediator masses $m_N \lesssim 5\, \textrm{MeV}$, we see a big improvement over the kaon decay constraints. Indeed in the $m_\delta = 0.1\,\textrm{MeV}$ scenario, we see an order of magnitude improvement across a decent range of mediator masses. There is also not a strong dependence on the neutrino masses. This suggest that even if the cosmological bounds on the total sum of neutrino masses tightens, these constraints will not weaken significantly. Furthermore, whilst there is a slight difference between the neutrino hierarchies due to the different couplings to the mass eigenstates, this is again not significant. Hence, we are not sensitive to future revelations from the neutrino sector.

In addition, we have also neglected additional processes that would decrease the mean free path. These include $\nu\delta \rightarrow \nu\delta$. As discussed in Section \ref{sec:oneprocess}, we do believe that this will be a negligble contribution. Nonetheless, again the conclusions are insensitive to this. This is because any decrease in the mean free path would only \emph{strengthen} the constraints.

We now turn to comparing our constraints with those from cosmology, in particular it turns out that the contribution to the neutrino temperature in the early Universe due to the interaction of the thermal bath with these light dark matter candidates is significant.  Modification of the neutrino temperature affects the CMB, for example by changing the epoch of matter radiation equality.

\begin{figure}[t]
 \centering
 \includegraphics[width=.8\textwidth]{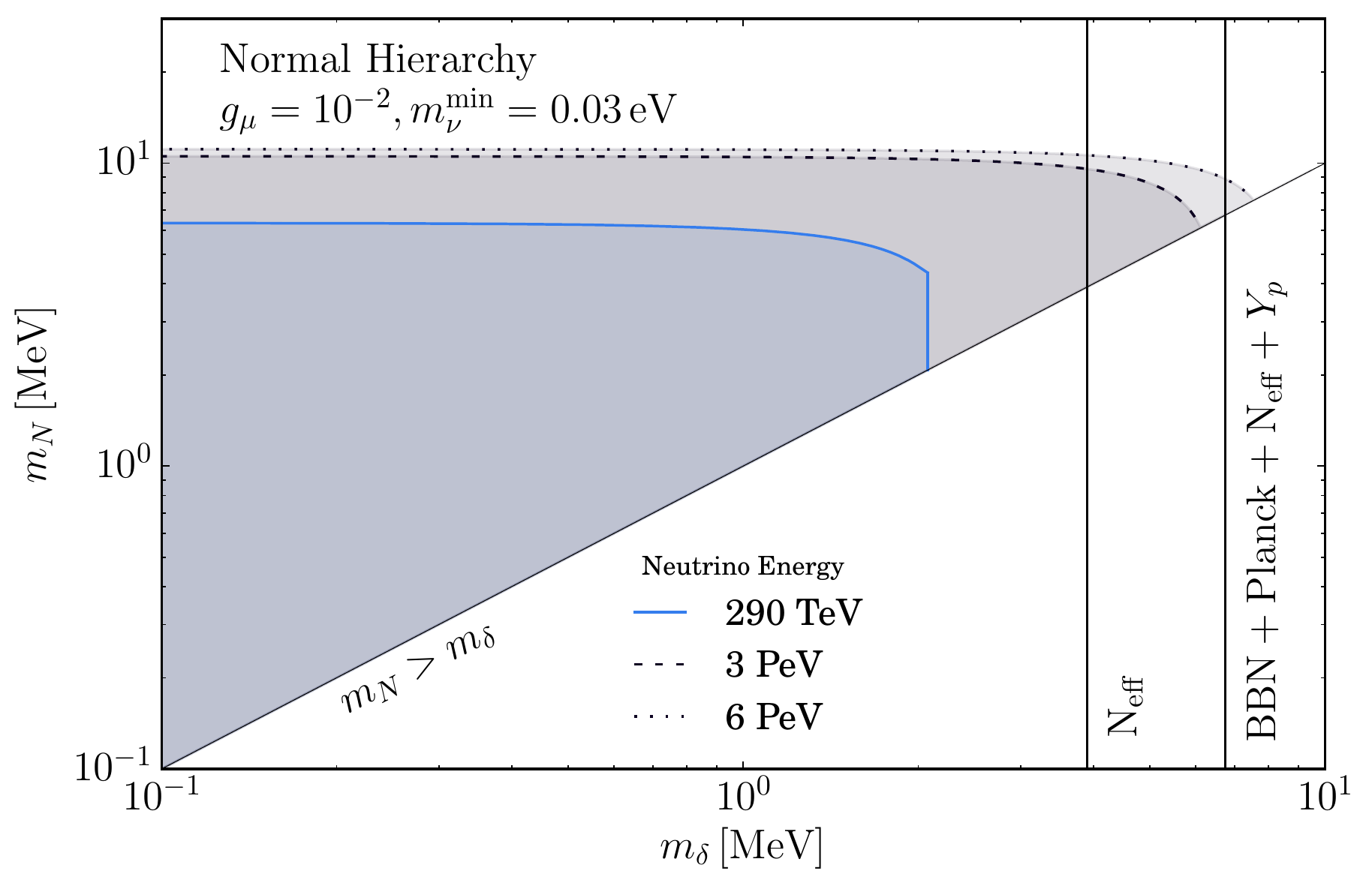}
 \caption{Expected constraints if higher energy neutrinos were detected from a source located at the same distance as TXS 0506+056. In this figure, we have taken $g_\mu = 10^{-2}$, $g_e = 0$, $g_\tau = 3 \times 10^{-1}$, and $m_\nu^{\mathrm{min}} = 0.03$ in the normal hierarchy. Shown vertically are CMB constraints \cite{Boehm} and CMB+BBN constraints \cite{Nollett2015}.}
 \label{fig:mpmn}
\end{figure}

In \cite{Boehm} the authors looked at the effect of a change in neutrino temperature on the CMB, while in \cite{Nollett2015}, the effect of a different temperature of neutrinos on both the CMB and BBN were also taken into account (see also \cite{Escudero2019}. These two studies lead to a stronger constraint on the mass of scalar dark matter coupled to the neutrino sector than the limits we are currently able to obtain from IceCube.  As an aside, it is interesting that two constraint from such radically different astrophysical environments lead to similar numbers. In Figure \ref{fig:mpmn} we plot these constraints neglecting any dependence on the mediator mass and we estimate what energy neutrino we would have to observe from TXS 0506+056 in order to get a stronger constraint. The CMB and BBN constraints on the model change the acceptable region where mass generation in this way is acceptable but leave enough parameter space open to not rule out the mechanism all together.

Taking a slightly bigger picture view, this shows that the exciting new data from IceCube can lead to new constraints on physics beyond the standard model, as shown also in \cite{Kelly}. Ultimately detecting neutrinos is difficult, and this analysis will only improve with more observations. As the IceCube experiment runs longer and longer, more regions of parameter space can be tested, as discussed in Section \ref{sec:cmbconstraints} and illustrated in Figure \ref{fig:mpmn}. It would also allow for a statistical treatment of the phenomenon which is simply not appropriate here with a single event.

To extend the analysis, we should consider the splitting of the mass eigenstates. In the case that there was a very large mass splitting, the constraints would weaken slightly as the centre of mass energy may not be sufficient to excite all the interaction modes. This will ultimately only lead to a maximum of a factor of $3$ difference in the mean free path, which is subdominant compared to changes in e.g. the coupling constants.

\section{Conclusions}\label{sec:conclusion}

To summarise, we have considered new constraints on an effective theory that links Dark Matter and neutrino masses. Within this framework, we used the IceCube event 170922A to obtain bounds on the interaction of neutrinos with MeV scalar dark matter. Our constraints are an order of magnitude improvement on the constraints which can be obtained by Kaon decay and are competitive with those coming from CMB and BBN, although not yet quite as strong.  We have calculated the energy of neutrinos that would need to be detected in order to beat the CMB and BBN constraint. Independent of the performance, it indicates the utility of IceCubce as a probe of fundamental physics. We also extended the analysis in \cite{Kelly} regarding the blazar dynamics to address some of the subtleties that arise in the higher energy neutrino regime. We argued that the various astrophysical papers \cite{Padovani2018, Padovani2019, Keivani2018} support the conclusion that the mean free path was a suitable comparison parameter in a fairly model independent fashion.

In terms of the assumptions contained in our analysis, we would like to ultimately relax the mass splitting assumption, although as noted above, we do not believe this would make a significant difference. For a very precise calculation, we should also include the effects of neutrino clustering for example.

Looking forward, the biggest improvement will be found once IceCube observes more ultra high energy astrophysical neutrinos from sources gigaparsecs away. The coincidence of the centre of mass energy and the scalar mass makes this sort of event the perfect setting to consider the effective model. Further work could then implement a fuller statistical analysis and provide confidence intervals on the bounds. One could also place similar constraints on the remaining portion of the models detailed in \cite{BoehmPascoli}. Moreover, one could look to investigate the amount of data required for clear discrimination between models. This would be a valuable step towards making this method of constraint a precise tool.

\acknowledgments{We would like to thank Miguel Escudero, Miguel Campos, Tevong You, and Diego Blas for their useful comments and discussions.  MF  is  funded  by  the  European  Research  Council  under  the  European Union's Horizon 2020 programme (ERC Grant Agreement no.648680 DARKHORIZONS).  In addition, the work of MF was supported partly by the STFC Grant ST/P000258/1.  JA is a recipient of an STFC quota studentship.}

\bibliography{main}
\newpage
\appendix
\section{Full Results}
\renewcommand{\thefigure}{\Alph{section}.\arabic{figure}}
\begin{figure}[h]
 \centering
  \includegraphics[width=.28\textwidth]{1.pdf}\quad
  \includegraphics[width=.28\textwidth]{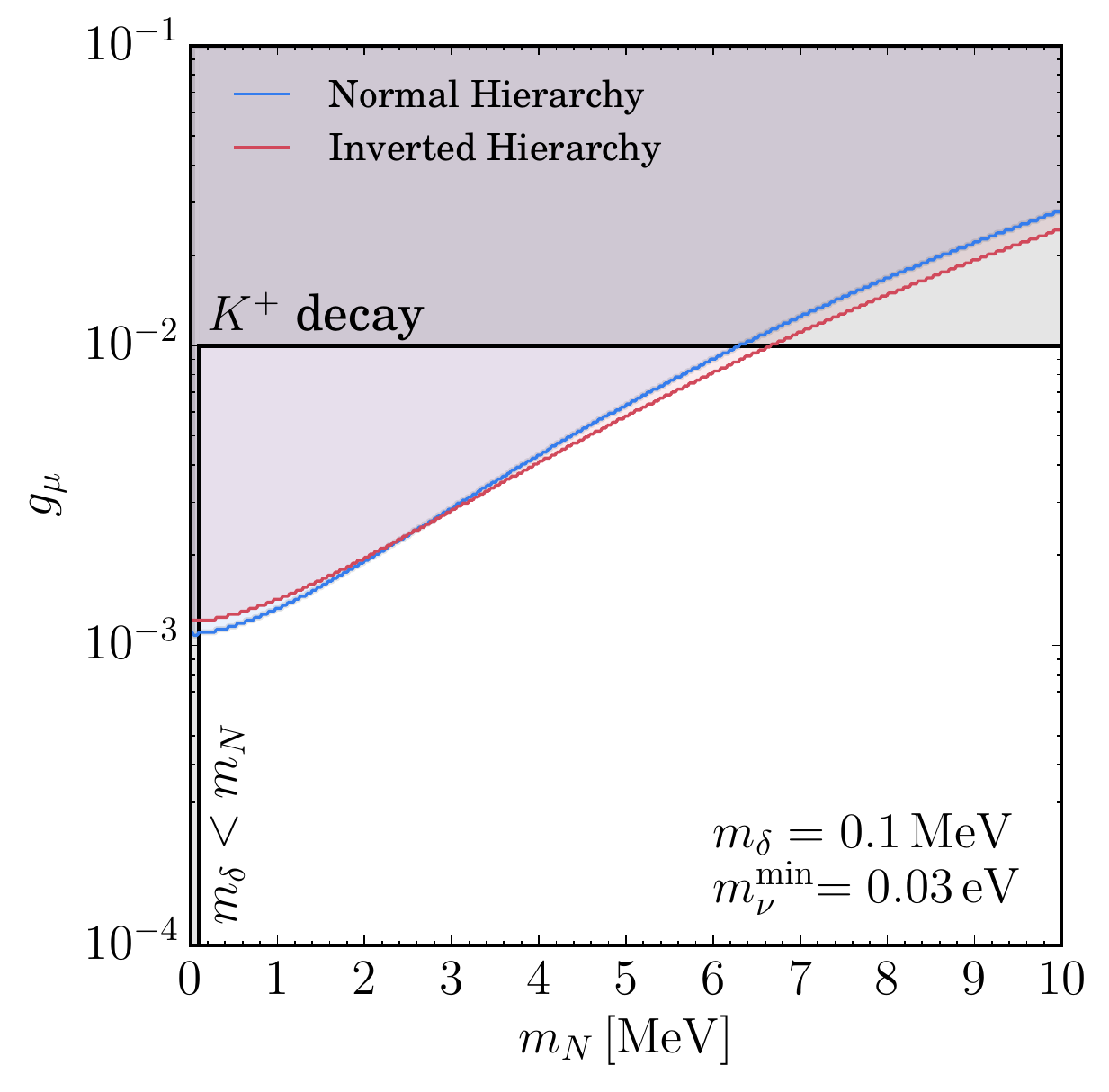}\quad
  \includegraphics[width=.28\textwidth]{3.pdf}

  \medskip

  \includegraphics[width=.28\textwidth]{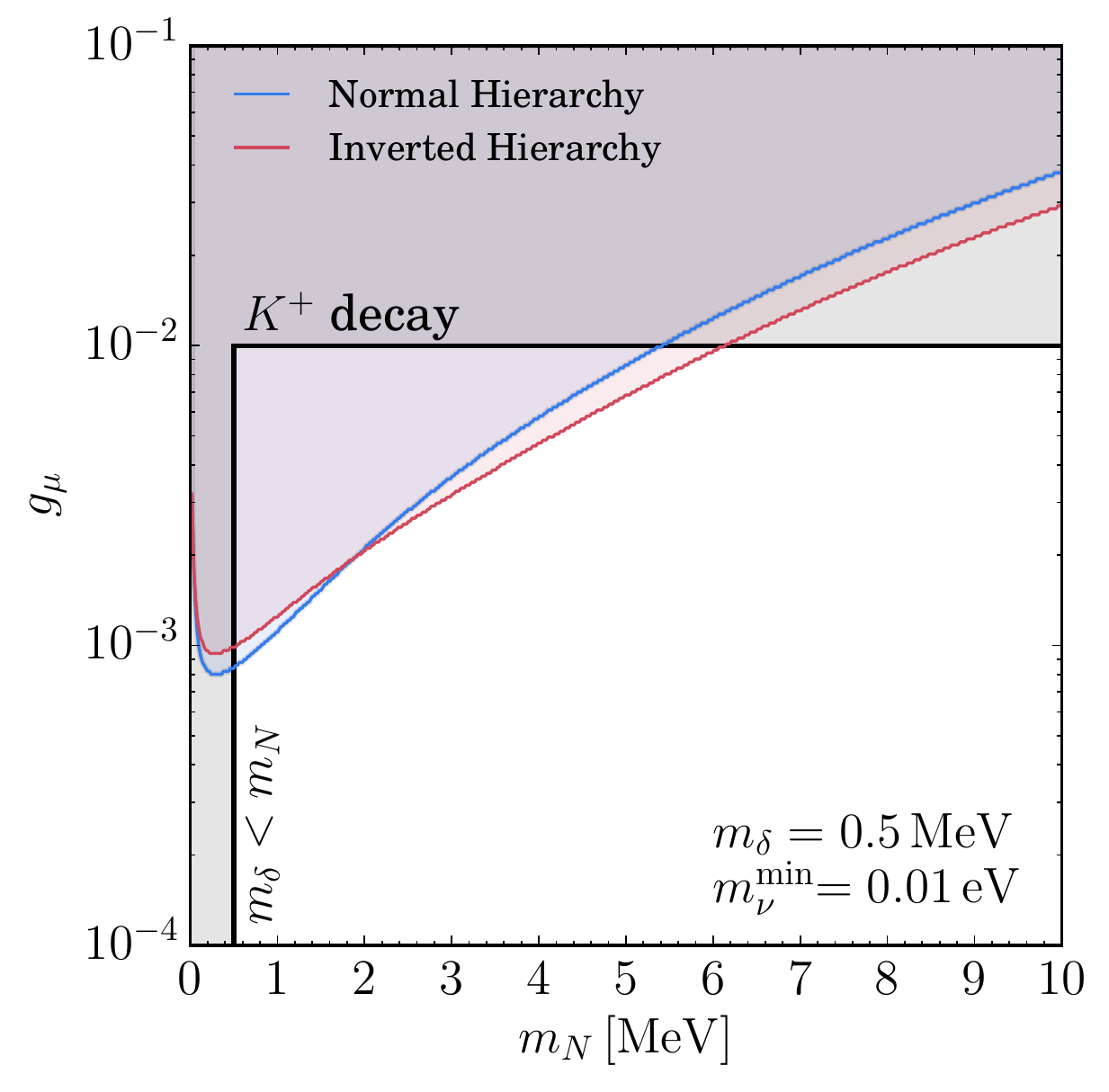}\quad
  \includegraphics[width=.28\textwidth]{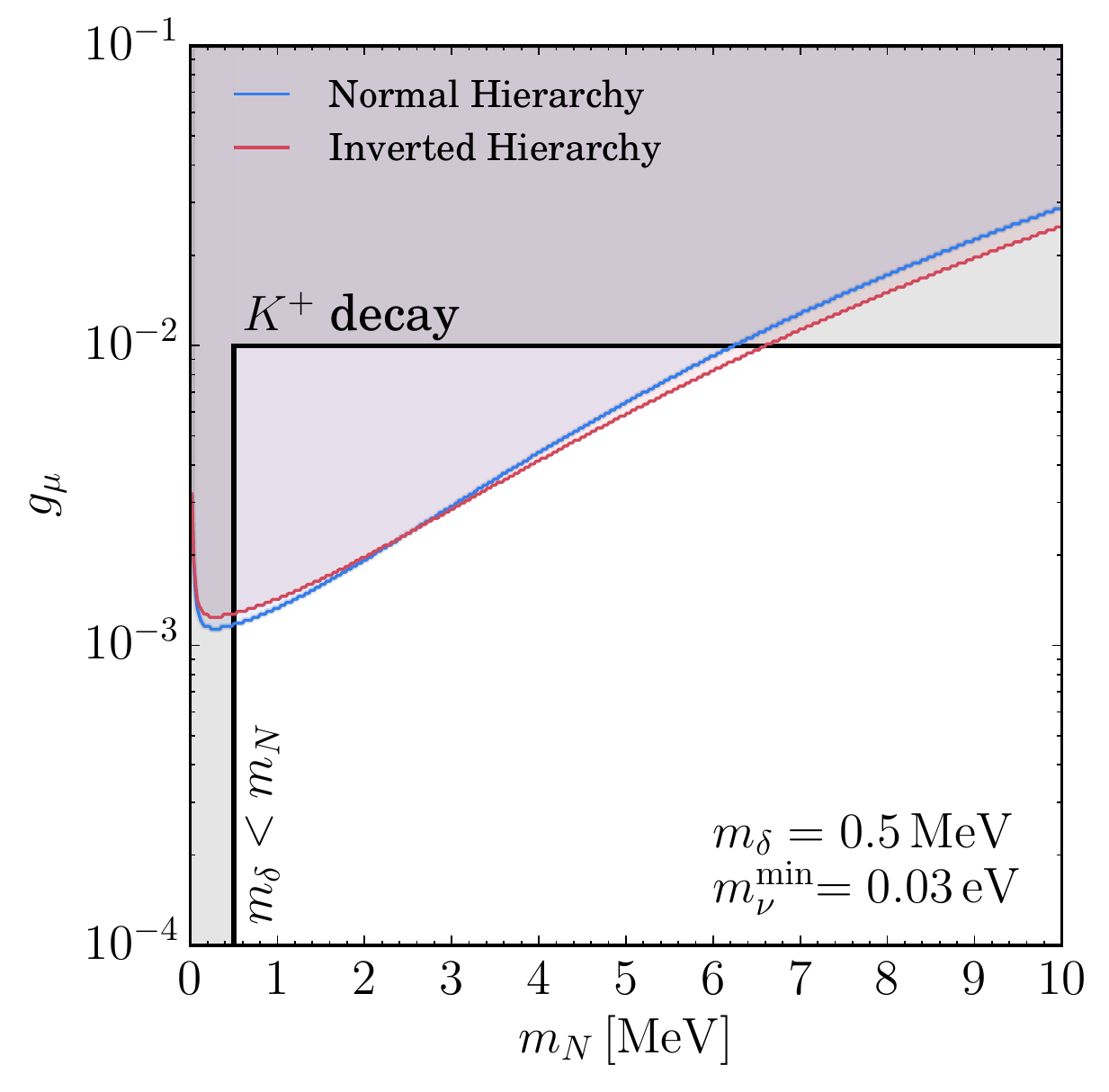}\quad
  \includegraphics[width=.28\textwidth]{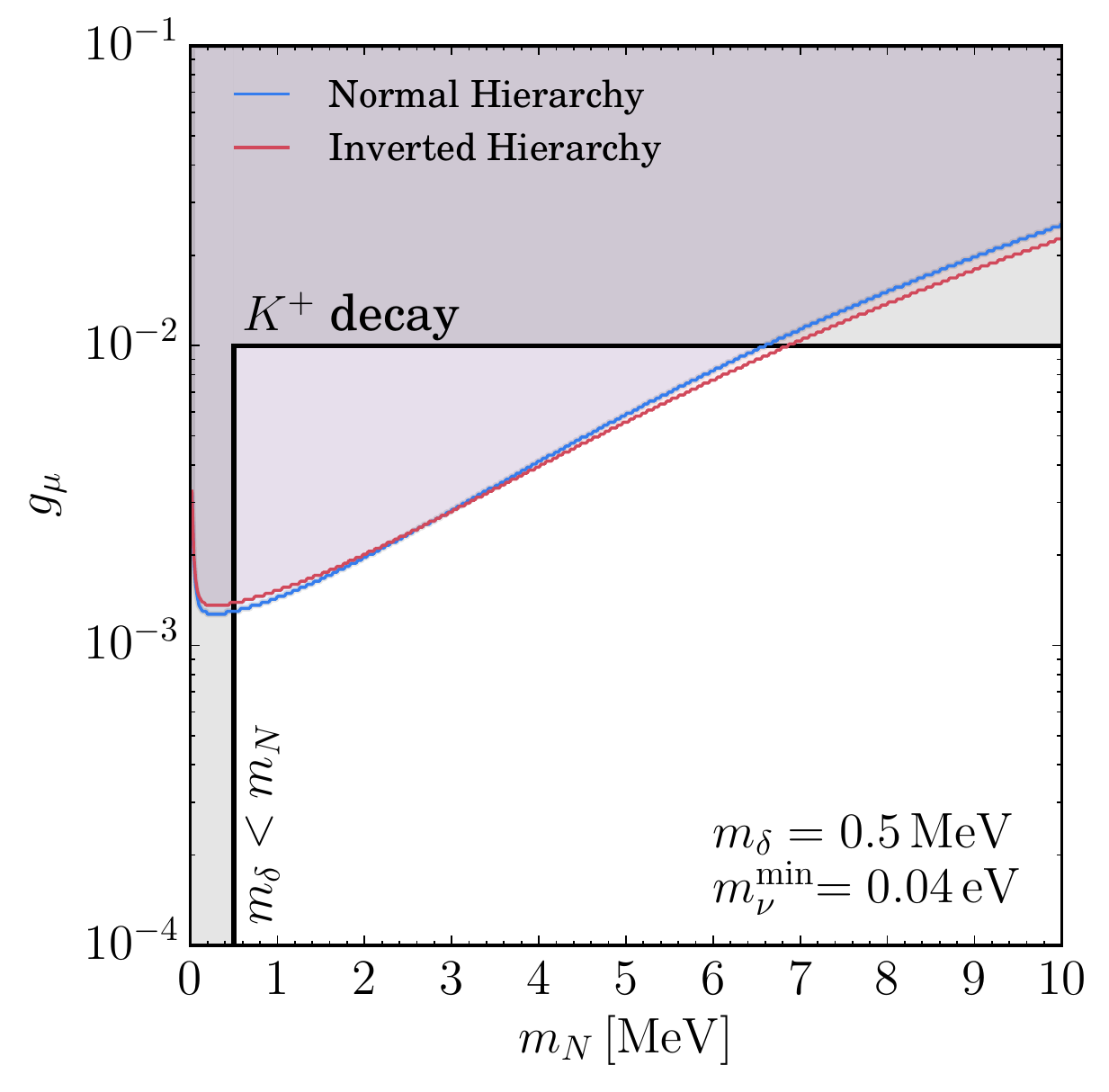}

  \medskip

  \includegraphics[width=.28\textwidth]{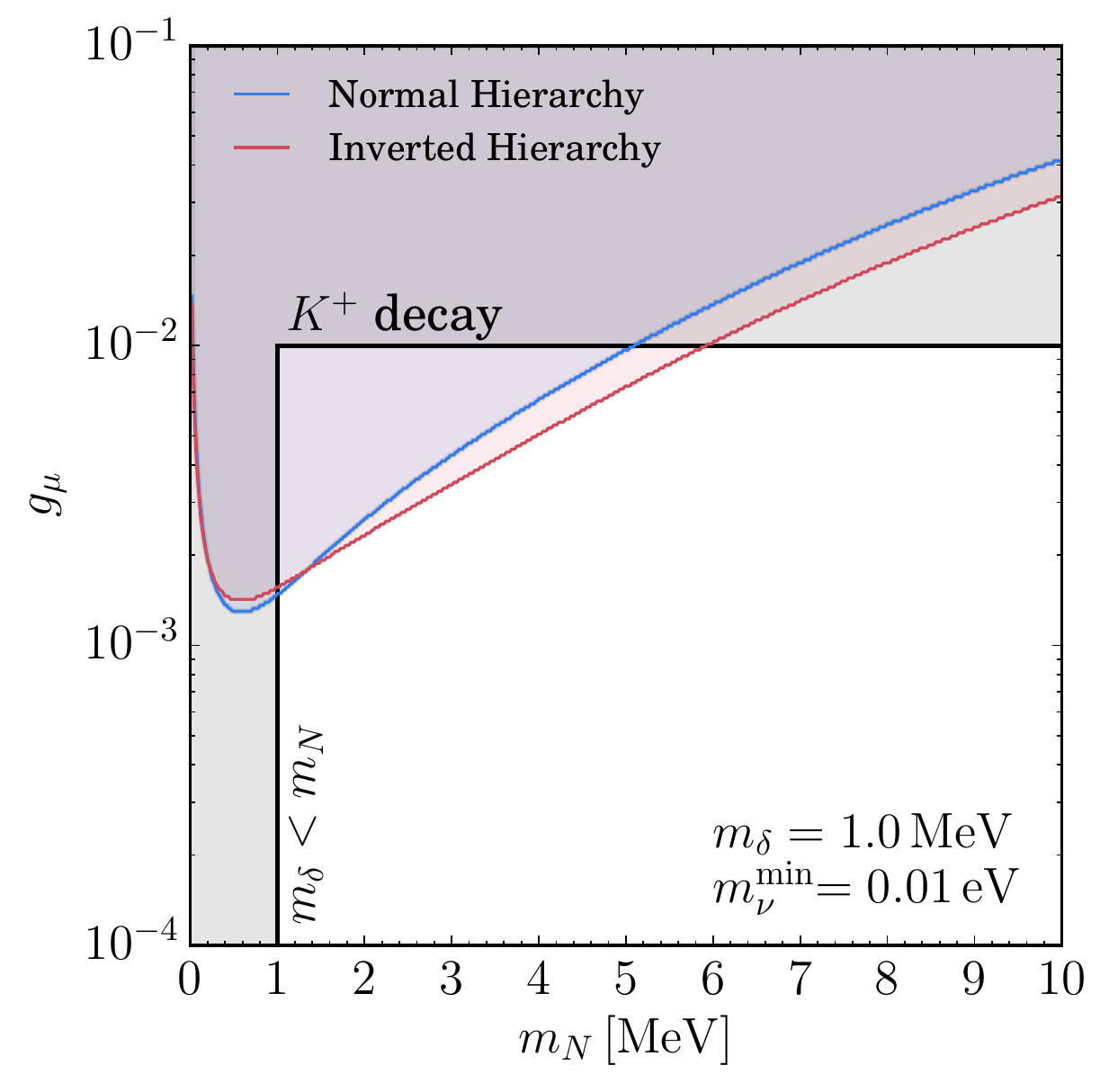}\quad
  \includegraphics[width=.28\textwidth]{8.pdf}\quad
  \includegraphics[width=.28\textwidth]{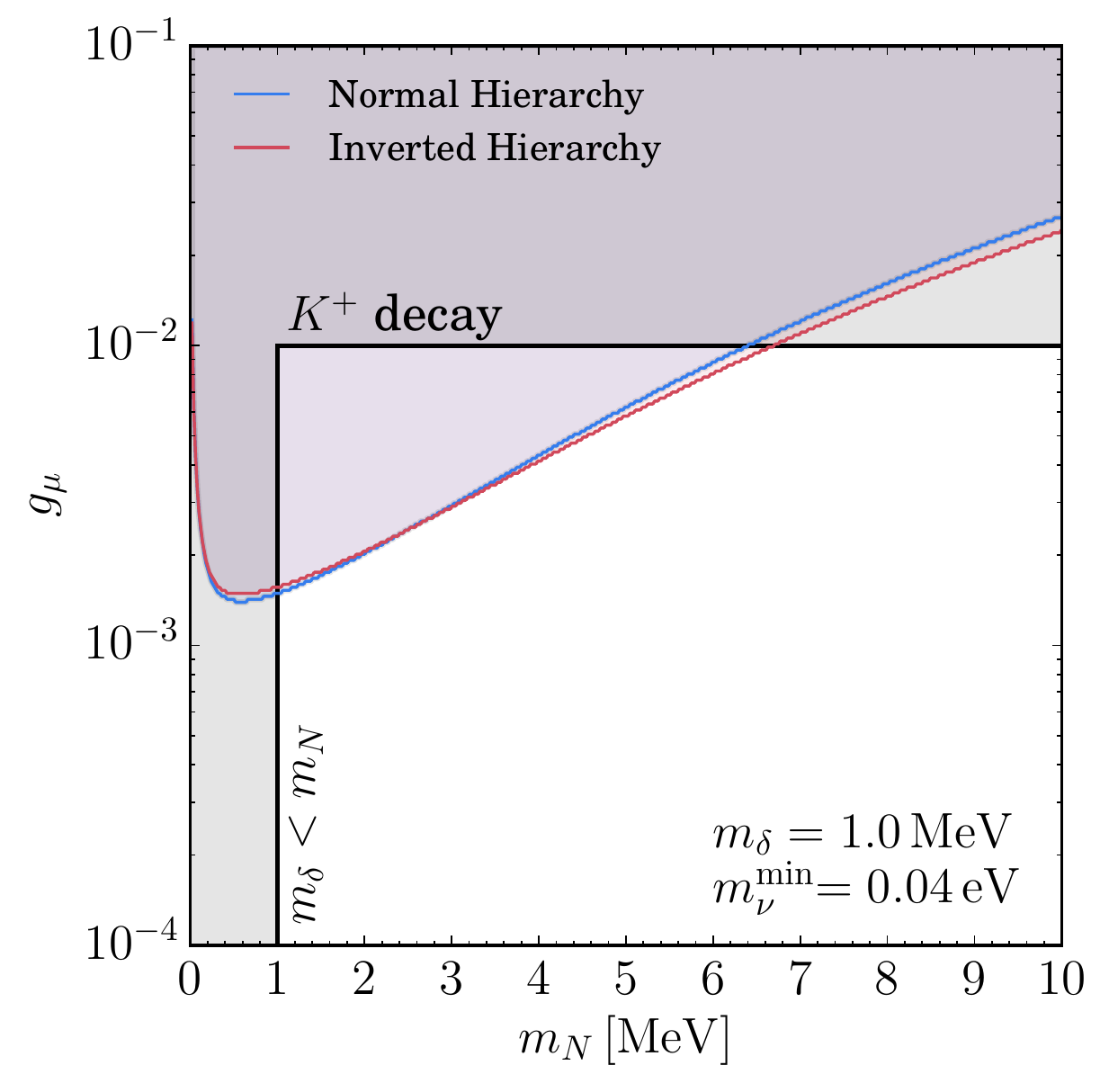}

  \medskip

  \includegraphics[width=.28\textwidth]{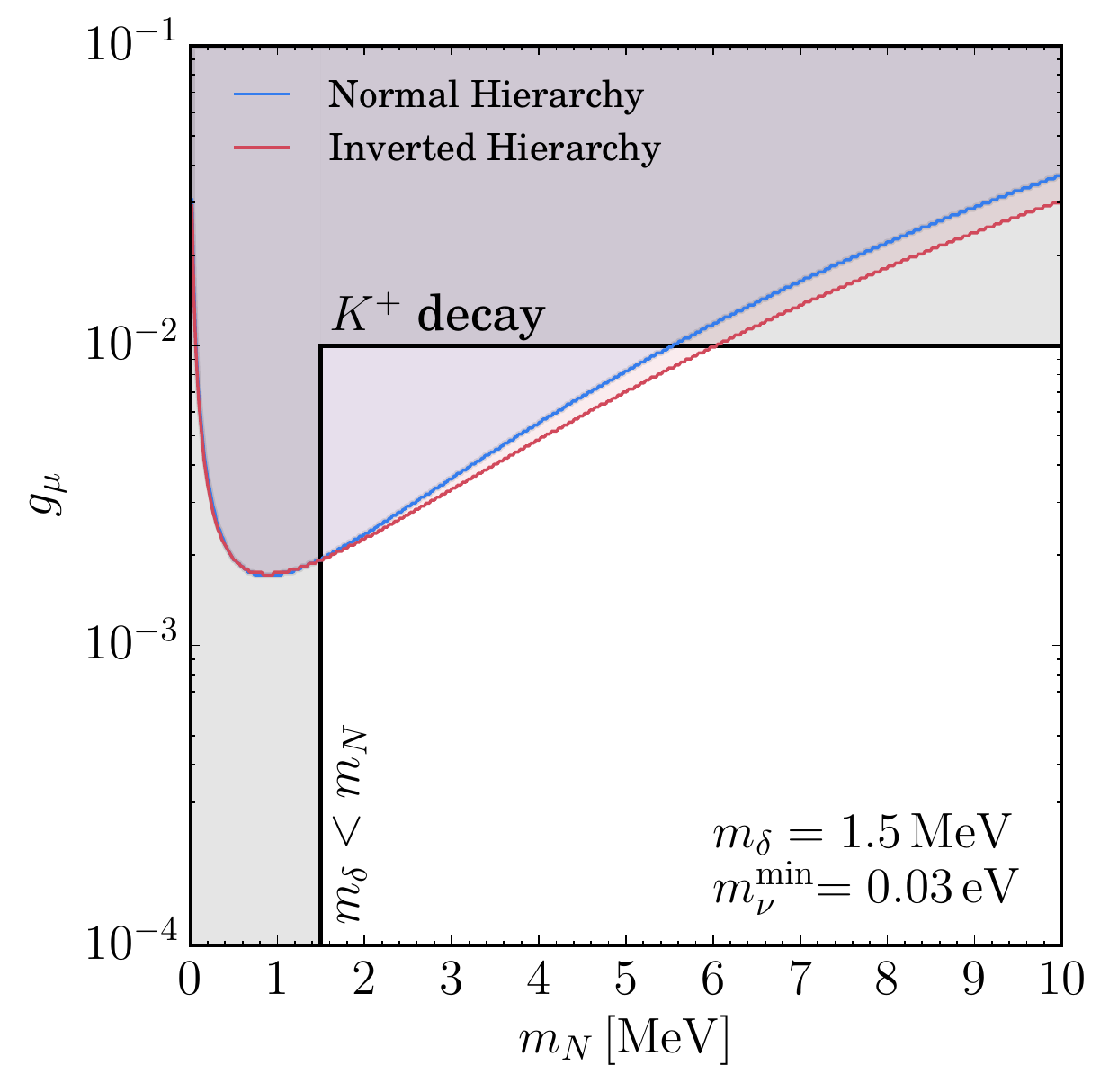}\quad
  \includegraphics[width=.28\textwidth]{11.pdf}
 \caption{Full results for the complex dark matter constraints.}
 \label{fig:appendix}
\end{figure}
\end{document}